\documentclass[12pt]{article}
\usepackage{epsfig}
\usepackage{graphicx}
\usepackage{a4}
\usepackage{latexsym}
\usepackage{cite}

\usepackage{color}
\usepackage{colordvi}
\usepackage{pslatex}
\usepackage[latin1]{inputenc}
\usepackage[T1]{fontenc}

\newcommand{\as}{\alpha_{\rm s}}
\newcommand{\ar}{a_{\rm s}}

\def\MSbar{\overline{\mathrm{MS}}}
\def\ep{\epsilon}
\def\z#1{{\zeta_{#1}}}
\def\ca{{C^{}_A}}
\def\cf{{C^{}_F}}
\def\nf{{n^{}_{\! f}}}
\def\nl{{n^{}_{\! l}}}
\def\nh{{n^{}_{\! h}}}
\def\ng{{n^{}_{\! g}}}

\def\tdC{{C}}
\def\tdG{{G}}
\def\calO{{\cal O}}

\def\H(#1){{\rm{H}}_{#1}}
\def\Hh(#1,#2){{\rm{H}}_{#1,#2}}
\def\Hhh(#1,#2,#3){{\rm{H}}_{#1,#2,#3}}
\def\Hhhh(#1,#2,#3,#4){{\rm{H}}_{#1,#2,#3,#4}}

\begin{document}
\setlength{\parskip}{0.2cm} \setlength{\baselineskip}{0.55cm}

\begin{titlepage}
\noindent
DESY 06-216 \hfill {\tt hep-ph/0612149}\\
SFB/CPP-06-53 \\
December 2006 \\
\vspace{2.0cm}
\begin{center}
\LARGE {\bf The singular behavior of massive QCD amplitudes} \\
\vspace{2.6cm}
\large
A. Mitov and S. Moch \\
\vspace{1.6cm}
\normalsize
{\it Deutsches Elektronensynchrotron DESY \\
\vspace{0.1cm}
Platanenallee 6, D--15738 Zeuthen, Germany}\\
\vfill
\large {\bf Abstract}
\vspace{-0.2cm}
\end{center}
We discuss the structure of infrared singularities in on-shell QCD
amplitudes with massive partons and present
a general factorization formula in the limit of small parton
masses. The factorization formula gives rise to an all-order
exponentiation of both, the soft poles in dimensional
regularization and the large collinear logarithms of the parton
masses. Moreover, it provides a universal relation between
any on-shell amplitude with massive external partons and its corresponding
massless amplitude. For the form factor of a heavy quark we present
explicit results including the fixed-order expansion up to three
loops in the small mass limit. For general scattering processes we
show how our constructive method applies to the computation of all
singularities as well as the constant (mass-independent) terms of a generic massive
$n$-parton QCD amplitude up to the next-to-next-to-leading order
corrections.
\\
\vspace{3.0cm}
\end{titlepage}

%
% -----------------------------------------------------------------------------
%
\section{Introduction}
\label{sec:intro}

Amplitudes for hard scattering processes in Quantum Chromodynamics (QCD)
are of basic importance both for theory and phenomenology
and predictions for these matrix elements
have to include higher-order quantum corrections.
They are mandatory for precision measurements of Standard Model parameters
and critical to the determination of backgrounds for new physics phenomena.
Many explicit computations of hard multi-parton processes do not only provide us
with a wealth of information but have also helped significantly in
understanding underlying principles such as factorization
or the universal structure of collinear and infrared singularities.

These singularities are particularly prominent for at least two reasons.
First of all, the independent knowledge of the universal limits when parton
momenta become collinear or a gluon momentum tends to zero
serves as a very strong check on any complete calculation.
Secondly, the calculation of finite cross-sections in QCD beyond
leading order has to combine consistently squared matrix elements
with different numbers of partons in the final state. In any such
formalism (see e.g.~\cite{Catani:1997vz}) the individual
contributions have to be suitably integrated over the available
phase space and are usually infrared divergent. At next-to-leading
order (NLO) in QCD the singular behavior of the corresponding amplitudes
with both massive and massless partons in the final state has been
extensively studied \cite{Catani:2000ef,Phaf:2001gc,Catani:2002hc}.

Research beyond NLO in the past years has been primarily focused
on the calculation of massless amplitudes
at next-to-next-to-leading order (NNLO), see for example~\cite{%
Bern:2003ck,DeFreitas:2004tk,Bern:2002tk,Glover:2003cm,Glover:2004si}
and numerous references therein.
The progress at NNLO and investigations of the singular behavior of
amplitudes at higher loops~\cite{Catani:1998bh} have significantly
contributed to our understanding of their general structure from the view point
of all-order resummations~\cite{Sterman:2002qn}.
This in turn leads to predictions for the soft and collinear
behavior of massless amplitudes at any order based on a small
number of perturbatively calculable anomalous dimensions.

For massive amplitudes however much less is known beyond NLO in QCD despite
the fact that NNLO precision predictions with massive quarks are clearly needed
in view of the data from present and the prospects of future high-energy colliders
(see Refs.~\cite{%
Czakon:2004wm,Czakon:2006pa,Passarino:2006gv} for related
progress).
Prominent examples of such measurements are for instance
the forward-backward asymmetry $A_{\rm FB}$ for inclusive heavy
quark production in $e^+e^-$-annihilation~\cite{:2005em,Bernreuther:2006vp},
and cross-sections for heavy flavor production and decays
at the Tevatron and the LHC (see e.g. Ref.~\cite{cms:2006tdr}).

The aim of this article is a first systematic investigation of the
structure of massive QCD amplitudes in singular limits beyond NLO.
To that end, we extend the studies of
Refs.~\cite{Catani:1998bh,Sterman:2002qn} to partonic scattering
processes including the presence of massive particles. The masses
of the latter screen the divergences of the massless amplitudes and
give rise to large logarithmically enhanced contributions of
Sudakov type~\cite{Sudakov:1954sw}, which dominate the high energy
behavior of the scattering amplitudes. It is precisely the
structure of these large logarithms together with soft
singularities appearing as poles in $(d-4)$ in $d$ dimensions, that we wish to
address here for a general non-Abelian ${\rm{SU}}(N)$-gauge theory
such as QCD.
Throughout the article we neglect power corrections in the parton masses $m$.

The outline of the paper is as follows. In
Section~\ref{sec:factorization} we recall the general framework for
the factorization of $n$-parton amplitudes in QCD and discuss its
modifications to incorporate massive partons.
As a result we derive an extremely simple universal multiplicative
relation between a massive amplitude in the small mass limit and
its massless version.
This is one main result of this paper.
The corresponding multiplicative factor (which we call $Z$)
can be linked to the QCD form factor of massive and massless partons.
Next in Section~\ref{sec:qcdffactor}, we specifically address
the resummation and exponentiation of the QCD form factor for heavy quarks,
which is our second main result.
On this basis we provide in Section~\ref{sec:rescoeff}
all resummation coefficients and new fixed-order expansions
of the massive form factor up to three loops.
For the resummation coefficients we observe striking relations between the massless
and the massive case.
In Section~\ref{sec:rescoeff} we also present explicit results
for the universal multiplicative factor $Z$
up to two loops and discuss its relation to the perturbative fragmentation function
of a heavy quark \cite{Mele:1990cw}.
We argue that our formalism represents the proper generalization of
Ref.~\cite{Mele:1990cw} at the level of amplitudes.
In Section~\ref{sec:applications} we demonstrate the predictive power
of the factorization ansatz for QCD amplitudes with examples from
$2 \to n$ scattering processes, such as hadronic $t \bar t$-production.
There we discuss the complete structure of the soft and collinear
singularities including the logarithmically enhanced terms to NNLO
in perturbative QCD.
We summarize in Section~\ref{sec:summary} and present some technical details in the
Appendix~\ref{sec:appA}.

%
% -----------------------------------------------------------------------------
%
\section{Factorization of QCD amplitudes}
\label{sec:factorization}

We are interested in a general $2 \to n$ scattering processes of partons $p_i$
\begin{equation}
\label{eq:QCDscattering}
{\rm p}: \qquad
p_1(k_1,m_1,c_1) + p_2(k_2,m_2,c_2) \:\:\rightarrow\:\:
p_3(k_3,m_3,c_3) + \dots + p_{n+2}(k_{n+2},m_{n+2},c_{n+2}) \, ,
\end{equation}
where $\{p_i\}$ denotes the set of partons (of specific flavors)
with associated momenta $\{ k_i \}$, masses $\{ m_i \}$ and
color quantum numbers $\{ c_i \}$. The latter are in the range $1 \dots N^2-1$
for particles in the adjoint (gluons) and $1 \dots N$ for particles
in the fundamental representation (quarks) of a ${\rm{SU}}(N)$-gauge theory.

The scattering amplitude ${\cal M}^{\rm[p]}$
for the process~(\ref{eq:QCDscattering}) is conveniently expressed
in a basis of color tensors $\left(c_I\right)_{\{c_i\}}$.
Following Ref.~\cite{Sterman:2002qn} we write
${\cal M}^{\rm[p]}$ as
\begin{eqnarray}
\label{eq:QCDamplitude}
{\cal M}^{\rm[p]}_{\{c_i\}}\left(\{ k_i \},{Q^2 \over \mu^2},\as(\mu^2),\ep \right)
&=&
{\cal M}^{\rm[p]}_{I}\left( \{ k_i \},{Q^2 \over \mu^2},\as(\mu^2),\ep \right)
\, \left(c_I\right)_{\{c_i\}} \\
&=& | {\cal M}_{\rm p} \rangle
\, ,
\nonumber
\end{eqnarray}
where $\mu$ is the renormalization scale, $Q$ the hard scale of the process
typically related to the center-of-mass energy, e.g. $Q = \sqrt{s}$ with
$s = (k_1 + k_2)^2$,
and $\ep$ the parameter of dimensional regularization, $d=4-2 \ep$.
The amplitude $| {\cal M}_{\rm p} \rangle $ is a vector in the space of
color tensors $c_I$ with summation over $I$ being understood.
We consider ${\cal M}^{\rm[p]}$ at fixed values of the
external parton momenta $k_i$, thus $k_i^2 = m_i^2$ and
especially $k_i^2 = 0$ for massless partons.
Any additional explicit dependence on the parton masses $m_i$ in
Eq.~(\ref{eq:QCDamplitude}) is suppressed.

Let us start by recalling that on-shell amplitudes for {\it massless} partonic processes
in $d=4-2 \ep$ dimensions can be factorized into products of functions
${\cal J}_0^{\rm[p]}$, ${\cal S}_0^{\rm[p]}$ and ${\cal H}^{\rm[p]}$.
These functions are called jet, soft and hard functions and
are known to organize the contributions of various momentum regions
relevant to the structure of the singularities in the scattering amplitude.
Following Refs.~\cite{Catani:1998bh,Sterman:2002qn} we can write
\begin{eqnarray}
\label{eq:QCDfacamplitude-zero}
| {\cal M}_{\rm p} \rangle
&=&
{\cal J}_0^{\rm[p]}\left({Q^2 \over \mu^2},\as(\mu^2),\ep \right)
{\cal S}_0^{\rm[p]}\left(\{ k_i \},{Q^2 \over \mu^2},\as(\mu^2),\ep \right)
| {\cal H}_{\rm p} \rangle
\, .
\end{eqnarray}
The short-distance dynamics of the hard scattering is described by
${\cal H}_{\rm p}$, which is infrared finite.
Analogous to the decomposition in Eq.~(\ref{eq:QCDamplitude})
$| {\cal H}_{\rm p} \rangle $ is a vector in the space of color tensors $c_I$.
Coherent soft radiation arising from the overall color flow is summarized by
${\cal S}_0^{\rm[p]}$, where we also use matrix notation suppressing
the color indices.
The function ${\cal J}_0^{\rm[p]}$ depends only on the external partons and
collects all collinearly sensitive contributions.
It is otherwise independent of the color flow.

Given the factorization formula~(\ref{eq:QCDfacamplitude-zero})
one can then organize the singularity structure of
any {\it massless} QCD amplitude.
After the usual ultraviolet renormalization is performed,
these singularities generally consist of two types, soft and collinear.
Being of infrared origin, they
are related to the emission of gluons with vanishing energy and
to collinear parton radiation off massless hard partons, respectively.
In this way all soft and collinear singularities in massless amplitudes
are regularized and appear as explicit poles in $\ep$ as indicated
in Eq.~(\ref{eq:QCDamplitude}).
Typically two powers of $1/\ep$ are generated per loop.

When masses are introduced the picture described above gets modified.
In QCD, which has only massless gauge bosons, the soft singularities remain
as single poles in $\ep$ while some of the collinear singularities are now
screened by the mass $m$ of the heavy fields.
Nevertheless, in presence of masses, we speak of {\it quasi-collinear}
singularities~\cite{Catani:2000ef}
that exhibit logarithmic dependence on $m$.
To be specific, in the present paper we will consider
the small mass limit of massive QCD amplitudes ${\cal M}^{\rm[p]}$
such as in Eq.~(\ref{eq:QCDamplitude}).
Naturally, in this limit we require that all masses in the
amplitude are either zero or equal to a common value $m$
and much smaller than the characteristic hard scale $Q$ of the reaction.
Thus, in the limit $Q^2 \gg m^2$ we aim at organizing all poles in $\ep$ and
all powers of $\ln^k(m)$, $k \ge 0$, (including mass independent terms)
from the underlying factorization principles.

From an alternative point of view however, the differences between a massless
and a massive amplitude for a given physical process
can also be thought of as a mere change in the regularization scheme.
Here, the limit of small masses for any given amplitude may simply
be seen as an alternative to working in $d$-dimensions in order to regulate
the soft and/or collinear singularities.
Of course, gauge invariance has to be retained.
In this interpretation parton masses act as formal regulators and
massive amplitudes in the limit $m^2 \ll Q^2$ must share essential properties
with the corresponding massless amplitudes. Such arguments have been
previously used in Refs.~\cite{Glover:2001ev,Penin:2005kf,Penin:2005eh}
in the context of QED corrections to the Bhabha process.
Within QCD with $\nl$ light quarks and one heavy flavor,
this requires to properly account for the decoupling of the heavy quark.
We will further elaborate on this point below, in particular on the
relevant aspects of the decoupling theorem~\cite{Appelquist:1974tg}.

Our goal is the generalization of the infrared factorization
formula~(\ref{eq:QCDfacamplitude-zero}) of Refs.~\cite{Catani:1998bh,Sterman:2002qn}
to the case of {\it massive} partons.
To that end, we perform a similar factorization for the amplitude ${\cal M}^{\rm[p]}$
into products of functions ${\cal J}^{\rm[p]}$, ${\cal S}^{\rm[p]}$
and ${\cal H}^{\rm[p]}$.
In the presence of a hard scale $Q$ we can then write
for the partonic process~(\ref{eq:QCDscattering})
\begin{eqnarray}
\label{eq:QCDfacamplitude}
| {\cal M}_{\rm p} \rangle
&=&
{\cal J}^{\rm[p]}\left({Q^2 \over \mu^2},{m_i^2 \over \mu^2},\as(\mu^2),\ep \right)
{\cal S}^{\rm[p]}\left(\{ k_i \},{Q^2 \over \mu^2},\as(\mu^2),\ep \right)
| {\cal H}_{\rm p} \rangle
\, ,
\end{eqnarray}
where all non-trivial mass dependence enters in the functions
${\cal J}^{\rm[p]}$ and ${\cal S}^{\rm[p]}$ and we neglect in ${\cal H}^{\rm[p]}$
power suppressed terms in the parton masses $m$.
The jet function now summarizes all {\it quasi-collinear}
contributions from the external partons. It is therefore of the form
\begin{eqnarray}
\label{eq:jetfactor}
{\cal J}^{\rm[p]}\left({Q^2 \over \mu^2},{m_i^2 \over \mu^2},\as(\mu^2),\ep \right)
&=&
\prod\limits_{i=1}^{n+2} {\cal J}^{[i]}\left({Q^2 \over \mu^2},{m_i^2 \over \mu^2},\as(\mu^2),\ep \right)
\, ,
\end{eqnarray}
where ${\cal J}^{[i]}$ denotes the jet function of each external parton $p_i$.

We stress that the above factorization formula~(\ref{eq:QCDfacamplitude})
is designed to correctly reproduce the leading power in the hard scale $Q$.
Moreover, as the similarity between Eqs.~(\ref{eq:QCDfacamplitude-zero})
and (\ref{eq:QCDfacamplitude}) suggests,
the factorization is otherwise independent of details such as the partons
in reaction~(\ref{eq:QCDscattering}) being massless or massive.
However, Eq.~(\ref{eq:QCDfacamplitude}) still
contains ambiguities related to the separation of finite terms
in ${\cal J}^{\rm[p]}$, ${\cal S}^{\rm[p]}$ and ${\cal H}^{\rm[p]}$.
It also contains ambiguities related to sub-leading soft terms
in ${\cal J}^{\rm[p]}$ and ${\cal S}^{\rm[p]}$.
Following Ref.~\cite{Sterman:2002qn} we fix this remaining freedom completely
by demanding that
\begin{eqnarray}
\label{eq:jetfunction}
{\cal J}^{[i]}\left({Q^2 \over \mu^2},{m_i^2 \over \mu^2},\as(\mu^2),\ep\right)
&=&
\left(
{\cal F}^{[i]}\left({Q^2 \over \mu^2},{m_i^2 \over \mu^2},\as,\ep\right)
\right)^{1 \over 2}
\, , \quad\quad i=q,g \, ,
\end{eqnarray}
where the scalar function ${\cal F}^{[i]}$ denotes
the gauge invariant space-like form factor of a quark or gluon
to be discussed in detail in Section~\ref{sec:qcdffactor} below.
For the moment, suffice it to say that the function ${\cal F}^{[q]}$
is associated to the vertex $\gamma^{\,\ast}\!qq$ (or $\gamma^{\,\ast}\!q\bar{q}$),
of a photon $\gamma^{\,\ast}$ with virtuality $Q^2$,
and $q/\bar{q}$ an external quark$\,$/$\,$anti-quark of mass $m_q$.
Likewise, for a colored parton in the adjoint ${\rm{SU}}(N)$-representation,
the function ${\cal F}^{[g]}$ is either
obtained from the effective vertex $\phi gg$ of a scalar Higgs and two
massless gluons, or from the corresponding vertex with two gluinos ${\tilde g}$
of mass $m_{\tilde g}$.

The motivation for the choice made in Eq.~(\ref{eq:jetfunction})
above comes from the following consideration. Firstly, it
reproduces the collinear dynamics as desired and, moreover,
provides a specific prescription for the pure soft terms contained
in the jet function. Secondly, it guarantees that the jet factor
${\cal J}^{\rm[p]}$ remains {\it process-independent}, while all
process-dependent soft interference terms are entirely delegated to
the soft function ${\cal S}^{\rm[p]}$. We recall that the role of
parton masses is to simplify screen the collinear singularities.
Since the soft and hard functions ${\cal S}^{\rm[p]}$ and ${\cal
H}^{\rm[p]}$ are insensitive to these collinear dynamics, being the
same in the massless or the massive case (provided $Q^2 \gg m^2$),
logarithmically enhanced contributions of the type $\ln^k(m)$ are
contained solely within ${\cal J}^{\rm[p]}$. In other words, we
require the (massive) factorization formula~(\ref{eq:QCDfacamplitude})
to be valid for any amplitude.
Then it also holds for the form factors ${\cal F}^{[i]}$ in
Eq.~(\ref{eq:jetfunction}) itself, since these are the simplest
amplitudes to which Eq.~(\ref{eq:QCDfacamplitude}) can be applied
with ${\cal S}^{[ii \to 1]} = 1 $ and ${\cal H}^{[ii \to 1]} = 1$,
and this choice for ${\cal J}^{\rm[p]}$ is also consistent with the
corresponding massless case.

We also want to comment briefly on evolution and exponentiation.
In Eqs.~(\ref{eq:QCDfacamplitude-zero}) and (\ref{eq:QCDfacamplitude})
we have suppressed any additional scale dependence, which together with
the renormalization group properties gives rise to evolution equations
for ${\cal J}_0^{\rm[p]}$, ${\cal S}_0^{\rm[p]}$ and
${\cal J}^{\rm[p]}$, ${\cal S}^{\rm[p]}$.
The solution of those evolution equations leads to an all-order exponentiation
in terms of the corresponding anomalous dimensions,
which is well known for massless partons, see e.g.
Refs.~\cite{Sterman:2002qn,MertAybat:2006mz}.
In the case of massive partons,
the exponentiation of the jet function ${\cal J}^{[i]}$
(the form factor ${\cal F}^{[i]}$, respectively)
is discussed in detail in Section~\ref{sec:qcdffactor},
while we postpone the soft function ${\cal S}^{\rm[p]}$ and its solution as a
path-ordered exponential until Section~\ref{sec:applications}.

Finally, the factorization formula~(\ref{eq:QCDfacamplitude}) along
with our choice~(\ref{eq:jetfunction}) for the jet function lends
itself to an even more suggestive form for practical applications,
namely, as a direct relation between the massless and the massive
amplitude, ${\cal M}^{{\rm[p]},(m=0)}$ and ${\cal
M}^{{\rm[p]},(m)}$, for any given physical process. To that end, we
exploit the full predictive power of Eq.~(\ref{eq:QCDfacamplitude})
and derive the remarkably simple and suggestive relation
\begin{eqnarray}
\label{eq:Mm-M0}
{\lefteqn{
{\cal M}^{{\rm[p]},(m)}\left(\{ k_i \},{Q^2 \over \mu^2},\as(\mu^2),\ep \right)
\, = \,}} \\
&&
  \prod_{i\in\ \{{\rm all}\ {\rm legs}\}}\,
  \left(
    Z^{(m\vert0)}_{[i]}\left({m^2\over \mu^2},\as(\mu^2),\ep\right)
  \right)^{1 \over 2}\,
  \times\
{\cal M}^{{\rm[p]},(m=0)}\left(\{ k_i \},{Q^2 \over
\mu^2},\as(\mu^2),\ep \right) \, ,
\nonumber
\end{eqnarray}
which is the first main result of this paper.

We have suppressed the color indices in Eq.~(\ref{eq:Mm-M0}). As we
see, the massless amplitude ${\cal M}^{{\rm[p]},(m=0)}$ and its
massive analogue ${\cal M}^{{\rm[p]},(m)}$ in the small mass limit
$m^2 \ll Q^2$ are multiplicatively related by a universal function
$Z^{(m\vert 0)}$. This result is consistent with Ref.~\cite{Catani:2000ef}
(see Section~\ref{sec:applications} for the detailed comparison).
The function $Z^{(m\vert 0)}$ is process independent and can be viewed
as a sort of renormalization constant (or rather a constant relating two
different regularization schemes).
This relation can be used to predict any massive amplitude from the known massless
one, the latter being much easier to compute in practice.
Moreover, Eq.~(\ref{eq:Mm-M0}) includes not only the singular terms
in the massive amplitude but extends even to the constant contributions
(i.e. the mass-independent terms).

With Eq.~(\ref{eq:jetfunction}) defining the jet function, the
function $Z^{(m\vert0)}$ is given in terms of the respective form
factors,
\begin{equation}
\label{eq:Z}
Z^{(m\vert0)}_{[i]}\left({m^2 \over \mu^2},\as,\ep
\right) \, = \, {\cal F}^{[i]}\left({Q^2\over \mu^2},{m^2\over
\mu^2},\as,\ep \right) \left({\cal F}^{[i]}\left({Q^2\over
\mu^2},0,\as,\ep \right)\right)^{-1}
\, ,
\end{equation}
where the index $i$ denotes the (massive) parton and $\as$ is
evaluated at the scale $\mu^2$.
Eq.~(\ref{eq:Z}) explicitly demonstrates the
process-independence of the factor $Z^{(m\vert0)}$. While both the
massive and the massless form factors are functions of the
process-dependent scale $Q$, this dependence cancels in their
ratio leaving in the factor $Z^{(m\vert0)}$ only the ratio of
process-independent scales $\mu^2/m^2$.

Although Eq.~(\ref{eq:Z}) is valid in a more general setting, and
in particular through any perturbative order, we will restrict in
the following our attention to QCD amplitudes and in particular to
those with massive quarks. For this case we will present explicit
results for $Z^{(m\vert 0)}_{[q]}$ up to two loops in
Section~\ref{sec:rescoeff}. Applications of Eq.~(\ref{eq:Mm-M0})
will be presented in Section~\ref{sec:applications}.

\begin{figure}[tb]
  \centering
  \vspace*{5mm}
  \includegraphics[width=15.0cm]{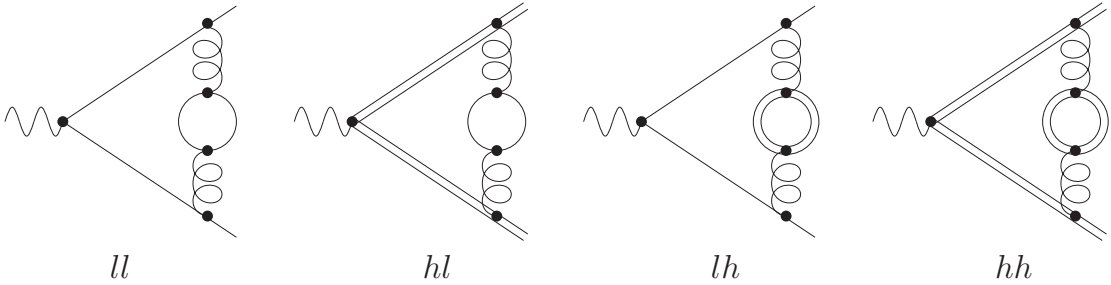}
  \caption{Feynman diagrams contributing to the
    vertex $\gamma^{\,\ast}\!qq$ as examples
    for the flavor classes
    $ll$, $hl$, $lh$ and $hh$ discussed in the text.
    Curly lines denote gluons, double straight lines quarks of mass $m$
    and single straight lines massless quarks.
  }
  \vspace*{5mm}
  \label{fig:flavorclass}
\end{figure}

Eqs.~(\ref{eq:Mm-M0}) and (\ref{eq:Z}) are
in addition subject to the following clarifications and qualifications.
First of all, the form factors entering in Eq.~(\ref{eq:Z}) for
$Z^{(m\vert 0)}_{[q]}$ are to be understood as being the form factors
in a theory with either $\nl+1$ massless quark flavors or
$\nl$ massless flavors and one heavy quark, respectively.
In both cases we have the same {\it total} number of flavors $\nf=\nl+1$.
Secondly, our approach of relating the large logarithms in $m$ to
{\it quasi-collinear} momentum regions requires external massive legs.
More precisely, we may define flavor classes,
according to the total number of heavy quark lines in an amplitude at a given order of perturbation theory.
In addition, the flavor classes distinguish for a given number of massive lines
whether the latter represent external legs or form closed internal loops.
For the form factor up to two loops,
we illustrate the various cases $ll$, $hl$, $lh$ and $hh$ in Fig.~\ref{fig:flavorclass}.
At tree level and one loop, we only have the pair of classes $ll$ and $hl$,
while from two loops onwards we also have the pair $lh$ and $hh$.
Both these pairs give rise to separate relations in Eq.~(\ref{eq:Z}).
Beyond two loops, yet new flavor classes can appear,
see e.g. Ref.~\cite{Vermaseren:2005qc}.
In fact a related discussion of this issue has already emerged in the literature
during the calculation of the NLO QCD corrections to the
three jet rate with massive quarks in electron-positron annihilation,
$e^+ e^- \to q {\bar q} X$, see e.g. Refs.~\cite{Brandenburg:1997pu,Nason:1997nw}.
It is also clear how to generalize the definition of flavor classes to
other types of colored heavy particles such as gluinos.

Let us finish this Section by pointing out another property of
Eq.~(\ref{eq:Mm-M0}). It is a standard textbook knowledge that the
two infrared regularizations of any one-loop QCD amplitude, either
with a quark mass or dimensionally, are related to each other as
follows
\begin{equation}
\label{eq:lnm->1/e}
\ln(m) \to {1\over \ep} + {\rm finite}\ {\rm terms}\ {\rm in}\ \ep
\, .
\end{equation}
Based on Eqs.~(\ref{eq:QCDfacamplitude}) and (\ref{eq:Mm-M0}),
we conclude in this paper that the proper generalization of Eq.~(\ref{eq:lnm->1/e})
beyond one loop is in the sense of process independent factorization.
The factor $Z^{(m\vert0)}$ in Eq.~(\ref{eq:Z}) is invertible and defines
the building block of proportionality to all orders in the strong coupling.

%
% -----------------------------------------------------------------------------
%
\section{The Sudakov form factor in QCD}
\label{sec:qcdffactor}

In the previous Section we have presented a factorization
that describes the singularity structure of QCD amplitudes both
in the massless case and in the limit of small masses $m^2 \ll Q^2$.
This factorization is valid through any perturbative order and
we have emphasized the central role of the form factor ${\cal F}^{[i]}$,
which specifically includes the QCD corrections.
Therefore, in this Section we want to focus on ${\cal F}^{[i]}$
and address the issue of its exponentiation.

To be precise we restrict the discussion here to ${\cal F}^{[q]}$ for the vertex
$\gamma^{\,\ast}\!q\bar{q}$
of a photon and an external quark-anti-quark pair, i.e.
to massive partons in the fundamental representation of the
${\rm{SU}}(N)$-gauge group.
Furthermore we confine ourselves to the case of one (heavy) external quark line
and no internal massive loops, which means we consider the flavor classes $ll$ and $hl$
(see Fig.~\ref{fig:flavorclass}).
We briefly comment on classes $lh$ and $hh$ at the end of this Section.
The gluon form factor ${\cal F}^{[g]}$ for all-massless partons
in the adjoint representation on the other hand,
which describes the vertex $\phi gg$ of a scalar Higgs and two gluons
is well known, see e.g. Ref.~\cite{Moch:2005tm}.
Also the necessary modifications to account for massive partons in the adjoint representation
such as gluinos in supersymmetric QCD have been worked out to one loop in Ref.~\cite{Catani:2000ef}.

Given a photon of virtuality $Q^2$
(we take space-like $q^2 = - Q^2 < 0$ throughout this Section)
the general expression for the vertex function $\Gamma_\mu$ reads
\begin{equation}
\label{eq:ffdef}
\Gamma_\mu(k_1,k_2)  \: = \:
{\rm i} e_{\rm q}\,
{\bar u}(k_1)\,
\left(
\gamma_{\mu\,}\, {\cal F}^{[q]}_1\! (Q^2,m^2,\as)
+
{1 \over 2 m}\,
\sigma_{\mu \nu\,} q^\nu \, {\cal F}^{[q]}_2\! (Q^2,m^2,\as)
\right) u(k_2)
\, .
\end{equation}
Here the external quark (anti-quark) of momentum $k_1$ ($k_2$)
is on-shell with $m$ denoting its mass and $e_{\rm q}$ its
charge, thus $k_1^2 = m^2$ (and $k_2^2 = m^2$).
The scalar functions ${\cal F}^{[q]}_1$ and ${\cal F}^{[q]}_2$
on the right-hand side are the space-like quark form factors,
which can be calculated order by order in the strong coupling
constant $\as$.
Results for the perturbative QCD corrections
to ${\cal F}^{[q]}_1$ in Eq.~(\ref{eq:ffdef})
are known through three loops in the massless on-shell case~\cite{%
Matsuura:1989sm,Moch:2005id,Gehrmann:2005pd,Moch:2005tm},
while the case of on-shell heavy quarks through two loops
has been considered in series of
papers~\cite{Bernreuther:2004ih,Bernreuther:2004th,Bernreuther:2005rw}.
${\cal F}^{[q]}_1$ and ${\cal F}^{[q]}_2$ are gauge invariant, but divergent
and in dimensional regularization with $d=4-2\ep$
these divergences show up as poles $\ep^{-k}$.
As we are concerned with the small mass limit $m^2 \ll Q^2$, we will in the
following mainly consider the pure vector-like form factor ${\cal F}^{[q]}_1$,
since ${\cal F}^{[q]}_2$ vanishes for massless quarks.
In the remainder we drop all indices and define
${\cal F} \equiv {\cal F}^{[q]}_1$.

The universality of soft and collinear radiation leads
on quite general grounds to an exponentiation of the respective
singular terms in the form factor,
be it poles in $\ep$ or large logarithms $\ln(m)$ of Sudakov type.
This has been well studied in the literature in various approaches~\cite{%
Collins:1980ih,Sen:1981sd,Sterman:1987aj,Korchemsky:1988hd,Catani:1989ne}.
Moreover, in the massless case explicit formulae have been given up to the
next-to-next-to-next-to-leading contributions~\cite{%
Magnea:1990zb,Magnea:2000ss,Moch:2005id}.
However, to the best of our knowledge, an equally valid exponentiated
representation for the massive form factor in dimensional regularization,
which holds beyond the leading contributions has still been lacking.
In this paper we present it for the first time.
In doing so we use two complementary derivations based
on evolution equations~\cite{Collins:1980ih}
and on inclusive partonic cross-sections~\cite{Catani:1989ne}.

Let us start with the former method and recall the
evolution equations for the form factor~\cite{Collins:1980ih}
\begin{equation}
\label{eq:ffdeq} -\ \mu^2 {\partial \over \partial \mu^2} \ln {\cal
F}\!\left({Q^2 \over \mu^2},{m^2 \over \mu^2},\as,\ep\right) \:  =
\:
  {1 \over 2} \: K\left({m^2 \over \mu^2},\as,\ep \right)
+ {1 \over 2} \: G\left({Q^2 \over \mu^2},\as,\ep \right) \, .
\end{equation}
The key input~\cite{Collins:1989bt} from QCD factorization are the dependence on the hard scale
$Q$ which rests entirely in the function $G$
and, to logarithmic accuracy, the separation of the mass dependence
in the function $K$.
Both functions, $G$ and $K$, are subject to renormalization
group equations~\cite{Collins:1980ih,Collins:1989bt},
\begin{equation}
\label{eq:KGdeq}
\mu^2 {d \over d \mu^2} G\left({Q^2 \over \mu^2},\as,\ep \right) \, = \,
- \lim_{m \to 0}
\mu^2 {d \over d \mu^2}
K\left({m^2 \over \mu^2},\as,\ep \right) \, = \,  A(\as) \, ,
\end{equation}
where we assume $\as = \as(\mu^2)$. Under renormalization group flow
both $G$ and $K$ are governed by the same
anomalous dimension $A$, because their sum is an invariant of the
renormalization group.
The anomalous dimension $A$ is well known for instance as the coefficient
of the $1/(1-x)_+$-contribution to the diagonal splitting functions or
alternatively as the anomalous dimension of a Wilson line with a
cusp~\cite{Korchemsky:1989si}.
Its power expansion in the strong coupling is currently known up to three
loops~\cite{Moch:2004pa,Vogt:2004mw}
and we use the convention (also employed for all other
expansions in $\as$ throughout this article)
\begin{equation}
\label{eq:abexp}
  A(\as)
\: = \: \sum\limits_{i=1}^{\infty} \left({\as \over 4\pi}\right)^i\, A_i
\: \equiv \: \sum\limits_{i=1}^{\infty} \left(\ar \right)^i\,  A_i\, ,
\end{equation}
where we have introduced the shorthand notation $\ar(\mu^2) \equiv
\alpha_s(\mu^2) / (4\pi)$ and similarly for the $d$-dimensional
coupling to be defined below. For later reference, we also mention
that we choose the $\MSbar$-scheme for the coupling constant
renormalization. The heavy mass $m$ on the other hand is always
taken to be the pole mass, thus the renormalization of $m$ imposes
the on-shell condition. We explicitly relate the bare
(unrenormalized) coupling $\as^{\rm{b}}$ to the renormalized
coupling $\as$ by
\begin{eqnarray}
\label{eq:alpha-s-renorm} \as^{\rm{b}} S_\epsilon \:
= \: Z_{\as}\, \as\, ,
\end{eqnarray}
where the renormalization constant $Z_{\as}$ in the $\MSbar$-scheme is given by
\begin{eqnarray}
\label{eq:alpha-Z}
Z_{\as} \: = \: 1 - {\beta_0 \over \epsilon} \ar\,
+ \left({\beta_0^2 \over \epsilon^2}
  - {1 \over 2} {\beta_1 \over \epsilon}\right) \ar^{\,2}\,
- \left({\beta_0^3 \over \epsilon^3}
   - {7 \over 6} {\beta_1 \beta_0 \over \epsilon^2}
   + {1 \over 3} {\beta_2 \over \epsilon}\right) \ar^{\,3}
\,
+ \dots
\, ,
\end{eqnarray}
and the bare expansion parameter is normalized as $\ar^{\rm{b}} =
\as^{\rm{b}} / (4\pi)\,$. For simplicity, we always set the ubiquitous factor
$S_\epsilon=(4 \pi)^\ep \exp(-\ep \gamma_{\rm E}) = 1$.

In Eq.~(\ref{eq:KGdeq}) all dependence on the infrared sector of the theory,
i.e. the structure of the singularities is described by the function $K$.
The function $G$, on the other hand, includes all dependence on the hard
scale $Q^2$ and is finite for $\ep \rightarrow 0$.
It is straight forward to solve the evolution equation~(\ref{eq:KGdeq}) for $G$.
Integration gives
\begin{eqnarray}
\label{eq:Gsol-massless} G\left({Q^2 \over \mu^2},\as,\ep \right) &
= & -\ G\left({\bar a}\left(Q^2,\ep \right) \right) -
\int\limits_{0}^{Q^2/\mu^2}\, {d \lambda \over \lambda}\ A({\bar
a}(\lambda \mu^2,\ep))\, ,
\end{eqnarray}
with the boundary condition $G({\bar a})$
to be derived by matching to fixed-order results for the form
factor.

Working in $d$-dimensions the solution for $G$ in
Eq.~(\ref{eq:Gsol-massless}) naturally depends on the
$d$-dimensional running coupling ${\bar a}(Q^2,\epsilon)$. The
latter can be expressed as a power series in the usual strong
coupling constant $\as(\mu^2)$ evaluated at a scale $\mu^2$. This
relation is now known through NNLO accuracy~\cite{Moch:2005id},
\begin{eqnarray}
\label{eq:arund} {\lefteqn{ \left({k^2 \over \mu^2}\right)^{\ep}
{\bar a}(k^2,\ep) \,=\, }}
\\
& & \mbox{}
    {\ar \over X} \biggl\{ 1 - \ep {\beta_1 \over \beta_0^2}
    {\ln X \over X} \biggr\}
  - {\ar^{\,2} \over X^2} \biggl\{ {\beta_1 \over \beta_0}
    (\ln X + Y) \biggr\}
  + {\ar^{\,3} \over X^3} \biggl\{
    {\beta_1^2 \over \beta_0^2} {3 \over 2} \ln^2 X
    \left( 1 + Y + {1 \over 4} Y^2 \right)
\nonumber\\
& & \mbox{}
    + {\beta_2 \over \beta_0} \ln X
      \left( {1 \over 6} (3 + Y) (1-X) - 1 - Y
      - {1 \over 3} Y^2 \right)
    \biggr\}
  + \calO \left( \ar^{\,4} \right) \, ,
\nonumber
\end{eqnarray}
which is consistent with the $\beta$-function in
$d$-dimensions~\cite{Magnea:2000ss,Contopanagos:1997nh}. Here we
have used $\ar = \ar(\mu^2)$, the obvious boundary condition ${\bar
a}(\mu^2,\ep) = \ar(\mu^2)$ and the abbreviations
\begin{equation}
\label{eq:XYpard}
  X \:=\: 1 - \ar {\beta_0 \over \ep}
  \left(\left({k^2 \over \mu^2}\right)^{-\ep} - 1\right)
  \:\: , \quad\quad
  Y \: = \: {\ep (1-X) \over \ar \beta_0} \, .
\end{equation}
The differential equation for $K$ in Eq.~(\ref{eq:KGdeq}) is similar
to the one for the function $G$ with the obvious difference that the
scale $Q$ is replaced by the mass $m$. The solution reads:
\begin{eqnarray}
\label{eq:K-massive} K\left({m^2 \over \mu^2},\as,\ep \right) & = &
-\ K\left({\bar a}\left(m^2,\ep \right) \right) +
\int\limits_{0}^{m^2/\mu^2}\, {d \lambda \over \lambda}\ A({\bar
a}(\lambda \mu^2,\ep))\, .
\end{eqnarray}

Combining the above results we obtain the solution of the evolution
equation~(\ref{eq:ffdeq}) for the form factor ${\cal F}$,
\begin{eqnarray}
\label{eq:massive-resummedff-explicit-form} {\lefteqn{ \ln {\cal
F}\!\left({Q^2 \over \mu^2},{m^2 \over \mu^2},\as,\ep\right) \, = \,
}}\\
&& - {1 \over 2} \int\limits_0^{Q^2/\mu^2}\, {d \xi \over \xi}
\Bigg\{
  G(\bar a(\xi \mu^2,\ep)) +
  \int\limits_{0}^\xi\, {d \lambda \over \lambda}\
  A({\bar a}(\lambda \mu^2,\ep))
\Bigg\}\nonumber\\
&& - {1 \over 2} \int\limits_0^{m^2/\mu^2}\, {d \xi \over \xi}
\Bigg\{
  K(\bar a(\xi \mu^2,\ep)) -
  \int\limits_{0}^\xi\, {d \lambda \over \lambda}\
  A({\bar a}(\lambda \mu^2,\ep))
\Bigg\}
 \quad \, , \nonumber
\end{eqnarray}
which satisfies the boundary condition ${\cal F}(0,0,\as,\ep) = 1$ and Eq.~(\ref{eq:ffdeq}).
Finally, we rearrange the above result as,
\begin{eqnarray}
\label{eq:massive-resummedff}
{\lefteqn{
\ln {\cal F}\!\left({Q^2 \over \mu^2},{m^2 \over \mu^2},\as,\ep\right)
\, = \,
}}\\
&&
- {1 \over 2} \int\limits_0^{Q^2/\mu^2}\, {d \xi \over \xi}
\Bigg\{
  G(\bar a(\xi \mu^2,\ep)) +
  K(\bar a(\xi \mu^2 m^2/Q^2,\ep)) +
  \int\limits_{\xi m^2/Q^2}^\xi\, {d \lambda \over \lambda}\
  A({\bar a}(\lambda \mu^2,\ep))
\Bigg\} \quad \, .\nonumber
\end{eqnarray}

Upon expansion of the $d$-dimensional coupling according to
Eq.~(\ref{eq:arund}), $\ln {\cal F}$ develops per power of $\ar$
double logarithms of $Q^2/m^2$ and single poles in $\ep$, which are
generated by the two integrations. To be specific, the single poles
are governed by the function $K$ in
Eq.~(\ref{eq:massive-resummedff}) and are generated only by the
outer $\xi$-integration. On the other hand, the inner
$\lambda$-integration over $A$ gives only rise to logarithms (as
long as the infrared cutoff is set by the heavy quark mass).
All quantities in Eq.~(\ref{eq:massive-resummedff})
are expressed in terms of the $d$-dimensional coupling $\bar a$.
In this way all integrations are regulated and no singularities
other than poles in $\ep$ arise.

Finally, after multiplying with a hard function $C$,
we are in a position to write down the exponential for the
complete massive form factor, which is our second main result,
\begin{eqnarray}
\label{eq:massFFexp}
{\lefteqn{
{\cal F}\left({Q^2 \over \mu^2},{m^2 \over \mu^2},\ar(\mu^2),\ep \right)
\, = \,
C({\bar a}(\mu^2,\ep),\ep)
\times }}
\\&&
\exp
\left[
- {1 \over 2} \int\limits_0^{Q^2/\mu^2}\, {d \xi \over \xi}
\Bigg\{
  G(\bar a(\xi \mu^2,\ep)) +
  K(\bar a(\xi \mu^2 m^2/Q^2,\ep)) +
  \int\limits_{\xi m^2/Q^2}^\xi\, {d \lambda \over \lambda}\
  A({\bar a}(\lambda \mu^2,\ep))
\Bigg\}
\right] \, , \nonumber
\end{eqnarray}
with all quantities on the right hand side being functions of ${\bar a}$ in $d$-dimensions.
Besides the known anomalous dimension $A$~\cite{Moch:2004pa,Vogt:2004mw}
all other functions $G$, $K$ and $C$ can be determined in a finite-order expansion.
This will be accomplished in Sec.~\ref{sec:rescoeff}.
Before doing so however, there is one more feature of Eq.~(\ref{eq:massFFexp})
which deserves some comment.

As is well known, for renormalization schemes used in QCD based on
dimensional regularization in the $\MSbar$-scheme,
the Appelquist-Carazzone decoupling theorem~\cite{Appelquist:1974tg}
does not hold true in its naive sense.
In a theory with $\nl$ light and $\nh$ heavy flavors
(thus $\nf = \nl + \nh$ for the total number of flavors)
the contributions of a heavy quark of mass $m$ to the
Green functions of gluons and light quarks
expressed in terms of the renormalized parameters of the full
theory do not exhibit the expected $1/m$ suppression.
The reason here is that the $\beta$-function governing the running
of the strong coupling constant $\as$ does not depend on any masses.
Neither do the anomalous dimensions describing the renormalization
scale dependence of all other parameters of the theory.
Rather, they exhibit discontinuities at the flavor thresholds,
which are controlled by so-called decoupling constants.

In the exponential expression Eq.~(\ref{eq:massFFexp}) for the form factor
we have used the standard $\MSbar$ coupling running with $\nl$ light flavors.
In order to compare Eq.~(\ref{eq:massFFexp})
or rather its expanded version to the fixed-order
calculations~\cite{Bernreuther:2004ih} of the massive form factor,
which also employ the $\MSbar$-scheme, but
a running coupling with a total number of flavors $\nf=\nl+1$,
one has to apply the decoupling relations.
The necessary decoupling constant for $\as$ at flavor thresholds is known to
$\calO(\alpha_s^3)$~\cite{Larin:1994va,Chetyrkin:1997sg,Chetyrkin:1997un}
(see also Ref.\cite{Chetyrkin:2000yt}).
To relate the two results, that is the expansion of Eq.~(\ref{eq:massFFexp})
on the one hand and the perturbative QCD corrections for the form factor
through two-loops~\cite{Bernreuther:2004ih} on the other, we use the
following relation for $\ar$,
\begin{eqnarray}
  \label{eq:adecd}
  \ar^{(\nl)} & = &
    \ar^{(\nf)}
    - {2 \over 3} L_{\mu, \ep} \biggl(\ar^{(\nf)}\biggr)^{\,2}
    + \biggl\{
    \left( {4 \over 9} - {\ep \over 3} (5 \ca + 3 \cf) \right) L_{\mu, \ep}^2
\\[1mm]
& & \mbox{}
    - {2 \over 3} (5 \ca + 3 \cf) L_{\mu, \ep}
    + {16 \over 9} \ca - {15 \over 2} \cf
    \biggr\} \biggl(\ar^{(\nf)}\biggr)^{\,3}
  + \calO \biggl( \biggl(\ar^{(\nf)}\biggr)^{\,4} \biggr) \, ,
\nonumber
\end{eqnarray}
where $\ar^{(\nl)}$ is the standard $\MSbar$ coupling for $\nl$
quark flavors expanded in terms of $\ar^{(\nf)}$ for $\nf=\nl+1$
flavors, both evaluated at the scale $\mu^2$. Eq.~(\ref{eq:adecd})
uses the pole-mass $m$.
The abbreviation $L_{\mu, \ep}$ denotes
\begin{equation}
\label{eq:Ldecd}
  L_{\mu, \ep} \:=\: {1 \over \ep}
  \left(\left({m^2 \over \mu^2}\right)^{-\ep} - 1\right)
  \, .
\end{equation}
Eq.~(\ref{eq:adecd}) is correct to NNLO and consistent with the standard
$\beta$-function in $d$-dimensions for all terms proportional to $L_{\mu, \ep}$.
For the constant terms at $\ar^3$ (i.e. those independent of $L_{\mu, \ep}$)
it is accurate up to terms of order $\ep$.
Eq.~(\ref{eq:adecd}) is to be inserted on the right hand side of Eq.~(\ref{eq:arund})
to decouple the heavy quark in the $d$-dimensional coupling.
Beyond one loop this generates in particular the correct
scale dependence to the accuracy required in Section~\ref{sec:rescoeff}.

Before moving on, we would like to discuss the exponentiation of
the massive form factor in Eq.~(\ref{eq:massive-resummedff}) from
a different perspective.
As announced above, our starting point here is the observation that
in sufficiently inclusive cross-sections, infrared singularities cancel
between real and virtual diagrams.
A suitable example for our purpose is the partonic cross-section
of inclusive deep-inelastic scattering (DIS) of a massive quark.
The purely virtual contributions to this partonic observable coincide
with the squared massive space-like form factor.

To extract the form factor, we first derive the all-order
exponentiation of the soft singularities of the cross-section for
the scattering of a massive quark $q$ off a virtual boson $V^*$, i.e.
$q+V^*\to q+X$.
To that end we follow the by-now standard methods for exponentiating
inclusive partonic cross-sections in Mellin $N$-space,
see e.g. Refs.~\cite{%
Sterman:1987aj,Catani:1989ne,Vogt:2000ci,Catani:2003zt,Moch:2005ba}.
Working in the eikonal approximation we obtain
\begin{equation}
\label{eq:DIS-plus}
\ln(\sigma(N,\as)) \,=\, \int\limits_0^1 dx \left(
{x^{N-1} -1 \over 1-x}\right) g(1-x,\as) \, ,
\end{equation}
where the function $g$ contains the powers of logarithms $\ln(1-x)$
at higher orders of $\as$ and $x$ is a kinematical variable related
to the Bjorken variable $x_B$, and to be specified below.

Secondly, we use the fact that the purely virtual diagrams exhibit
a simple $x$-dependence proportional to $\delta(1-x)$, i.e. in $N$-space
they contribute an $N$-independent factor.
Thus, working in the eikonal approximation
one can identify the contribution from the squared form factor
(to all orders in the strong coupling)
with the term "$-1$" in the factor $(x^{N-1} -1)$
in Eq.~(\ref{eq:DIS-plus}).
The complementary, $N$-dependent factor is entirely
related to real emission diagrams.
This way one can identify the logarithm of the form factor with the function
\begin{equation}
\label{eq:DIS-FF}
- {1\over 2}\int\limits_0^1 {dz \over z} g(z,\as) \, ,
\end{equation}
where $z=1-x$.

As it stands Eq.~(\ref{eq:DIS-FF}) is not well defined.
The reason is that it contains unregulated soft singularities.
Their appearance is not unexpected, since the factor
$(x^{N-1} -1)$ in Eq.~(\ref{eq:DIS-plus}) is constructed such
that it ensures the cancellation between the soft singularities
from the real and virtual corrections.
Moreover, it is precisely this cancellation that leads
to the appearance of the large distributions $[\ln^k(1-x)/(1-x)]_+$
(or large logarithms $\ln(N)$ in Mellin space) in the function $g$.
Therefore, if one removes the real emission contributions in
Eq.~(\ref{eq:DIS-plus}), one can no longer rely on the delicate
balance between real and virtual contributions to regularize the
soft singularities.
Clearly an alternative regularization of the latter is needed to render
Eq.~(\ref{eq:DIS-FF}) meaningful.

Since in this paper we are interested in regularizing the soft
divergences in the massive form factor (or in any other amplitude)
dimensionally, and in line with our previous discussion, we modify
Eq.~(\ref{eq:DIS-FF}) by replacing the usual coupling $\as$ with the
$d$-dimensional one $\bar a$ as defined in Eq.~(\ref{eq:arund}),
\begin{equation}
\label{eq:DIS-FF-fin}
\ln({\cal F}(\alpha_s)) \,=\, - {1\over 2}\int\limits_0^1 {dz \over z} g(z,\bar a) \, .
\end{equation}
We stress that the function $g$ in Eq.~(\ref{eq:DIS-FF-fin}) is the
same one that appears in the cross-section in
Eq.~(\ref{eq:DIS-plus}). The effect of the $d$-dimensional coupling
is rather transparent, as it supplies additional powers of the
factor $z^{-\ep}$, see e.g. the left hand side of Eq.~(\ref{eq:arund}),
which allows to regulate the $z$-integration
in Eq.~(\ref{eq:DIS-FF-fin}) in the limit $z \to 0$.

For the derivation of the required hard cross-section for the
process $q+V^*\to q+X$ we directly build on previous work on the
exponentiation of massive cross-sections at next-to-leading
logarithmic accuracy~\cite{Corcella:2003ib,Cacciari:2002re}, where
the light quark initiated process $q_l+V^*\to q+X$ was studied.
Since in this work we are interested in the corresponding process
initiated by heavy quarks, $q+V^*\to q+X$, one has to modify the
analysis of Ref.~\cite{Corcella:2003ib}. One possible option is to
repeat the considerations of that reference keeping a non-vanishing
mass for the initial state quark. However, a much simpler
alternative is to express the coefficient function for the
$q$-initiated process as a convolution of a perturbative
distribution function for the initial-state heavy
quark $q$ and the coefficient function for the process $q_l+V^*\to
q+X$ both evaluated at a common factorization scale $\mu_F$. Since
we are interested only in contributions that are enhanced in the
soft limit and suppress power corrections with the quark mass $m$,
only the $q \to q$ component of this distribution
function is required. Moreover in the soft limit this function with
space-like kinematics coincides with its time-like counterpart (see
e.g. Ref.~\cite{Gardi:2005yi}). All components of the time-like
perturbative fragmentation function $D$ are known through two loops
and can be found in Refs.~\cite{Melnikov:2004bm,Mitov:2004du}.

The exponential structure in the soft limit of the perturbative fragmentation
function $D$ of a heavy quark~\cite{Mele:1990cw} is well
understood~\cite{Korchemsky:1992xv,Cacciari:2001cw,Gardi:2005yi}.
In Mellin-$N$ space we have
\begin{eqnarray}
\label{eq:res-Dini}
\ln( D(N) ) = \int\limits_0^1 dx\ {x^{N-1}-1\over 1-x}\
\Bigg\{\  H\left(\as((1-x)^2m^2)\right)\ +\
\int\limits_{(1-x)^2m^2}^{\mu_F^2}\ {dk^2\over k^2}\ A\left(\as(k^2)\right) \Bigg\} \, ,
\end{eqnarray}
where the anomalous dimension $A$ is the same as the one appearing
in Eq.~(\ref{eq:abexp}) while $H$ is a new function.

The exponentiation of the coefficient function for the process
$q_l+V^*\to q+X$ was clarified in Ref.~\cite{Corcella:2003ib}.
With the same anomalous dimension $A$ and a new function $S$
the result reads,
\begin{eqnarray}
\label{eq:res-sigma0}
\ln( \sigma_{q_l \to q}(N)) =
\int\limits_0^1 dx\ {x^{N-1}-1\over 1-x}\
\Bigg\{\
S\left(\as((1-x)^2M^2)\right)\ +\
\int\limits_{\mu_F^2}^{(1-x)^2M^2}\ {dk^2\over k^2}\ A\left(\as(k^2)\right)
\Bigg\} \, ,
\end{eqnarray}
where in the limit $m^2\ll Q^2$ the scale $M$ equals $M^2 = Q^4/m^2$.

Let us briefly recall a few basic facts~\cite{Corcella:2003ib} about
the derivation of Eq.~(\ref{eq:res-sigma0}).
The variable $x$, $0\leq x\leq 1$, is the rescaled Bjorken variable
$x = (1+m^2/Q^2)x_B$.
The upper limit of the $k^2$-integration follows
from kinematics and in the center-of-mass frame it is determined from
the light quark energy $E:\ k^2\leq 4E^2(1-x)^2$.
Moreover one has
\begin{eqnarray}
\label{eq:PFF-kinematic}
2E\ \simeq\ {Q^2+m^2 \over \sqrt{(1-x)Q^2+m^2}}
\, .
\end{eqnarray}
Since we are working in the soft limit $(1-x)\to 0$, it is obvious
that in the massive case Eq.~(\ref{eq:PFF-kinematic}) leads exactly
to the scale $M$ in Eq.~(\ref{eq:res-sigma0}), while for $m=0$ it
reduces to the well known expression of the massless
case~\cite{Catani:1989ne}.

Convoluting Eqs.~(\ref{eq:res-Dini}) and (\ref{eq:res-sigma0}) we
obtain the desired coefficient function for the sub-process
$q+V^*\to q+X$ in the soft limit. One can see that the dependence
on the factorization scale drops out as it should. Following the
procedure outlined around
Eqs.~(\ref{eq:DIS-FF}), (\ref{eq:DIS-FF-fin}) above, we finally obtain the
Sudakov exponent for the massive form factor:
\begin{eqnarray}
\label{eq:res-FF}
\Delta_F\  =\ - {1\over 2} \int\limits_0^1 {dz\over z}\
\Bigg\{\
S\left({\bar a}(z^2M^2,\ep)\right) \ +\
H\left({\bar a}(z^2m^2,\ep)\right) \ +\
\int\limits_{z^2m^2}^{z^2M^2}\ {dk^2\over k^2}\ A\left({\bar a}(k^2,\ep)\right)
\Bigg\} \, .
\end{eqnarray}
The complete form factor is obtained by multiplying the above
exponent with a hard function $H_F$:
\begin{eqnarray}
\label{eq:res-FF-fin}
{\cal F}\!\left({Q^2 \over \mu^2},{m^2 \over \mu^2},\as,\ep\right)
&=&
H_F(Q^2,m^2,{\bar a}(\mu^2,\ep),\ep)\ \exp\bigl\{\Delta_F\bigr\} \, .
\end{eqnarray}
Here all functions $A$, $H$, $S$ and $H_F$ have perturbative expansions
analogous to Eq.~(\ref{eq:abexp}).
They can be obtained to a given order in $\as$ by matching for instance
to the full calculation for ${\cal F}$.
In addition, independent information on $H$ and $S$ arises also
with the help of
Eqs.~(\ref{eq:res-Dini}) or (\ref{eq:res-sigma0}) from the calculation
of the perturbative fragmentation function $D$ or the hard
partonic light-to-heavy DIS cross-section $\sigma_{q_l \to q}$.

The hard function $H_F$ has an expansion in $\ep$
but is finite in the limit $\ep\to 0$, since all soft poles are
collected in the exponent $\Delta_F$.
To completely define the hard function $H_F$
one has again to specify the definition of the coupling $\as$
appearing in Eqs.~(\ref{eq:res-FF}) and (\ref{eq:res-FF-fin}).
This is the usual $\MSbar$ coupling defined in Eq.~(\ref{eq:alpha-s-renorm})
but running {\it only} with the number of light flavors $\nl$.
The same number of flavors appears also in the anomalous dimensions,
see e.g. Refs.~\cite{Gardi:2005yi,Melnikov:2004bm}.
Thus, in order to compare Eq.~(\ref{eq:res-FF-fin})
to the fixed-order calculation available in, say Ref.~\cite{Bernreuther:2004ih},
with a coupling constant $\alpha_s$ for $\nf=\nl+1$ flavors,
we again have to apply the decoupling relations~\cite{%
Appelquist:1974tg,Larin:1994va,Chetyrkin:1997sg,Chetyrkin:1997un}
in the form of Eq.~(\ref{eq:adecd}).

A comparison to the exact two-loop calculation of the vector
form factor~\cite{Bernreuther:2004ih} shows that Eq.~(\ref{eq:res-FF-fin})
correctly predicts all soft terms $\sim 1/\ep^k~,~k\geq 1$
including their logarithmic mass dependence,
while it does not control the powers of $\sim\ln^k(m)$ at order $\ep^0$
which are of collinear origin.
From the viewpoint of the exponentiation of {\it soft} singularities
these latter logarithms must be included in the hard function $H_F$.

However, at the same time one expects that all pure collinear
logarithms exponentiate as well. This feature is unrelated to the
soft-gluon exponentiation discussed above but rather to the
standard parton evolution equations (DGLAP). Here we recall the
analysis of Ref.~\cite{Catani:1989ne} where the effect of collinear
radiation in the outgoing jet results in modifications of the naive
eikonal exponentiation. The additional collinear contributions in
the final-state are taken into account by constructing a DGLAP-like
evolution equation for the corresponding jet function.
The latter, in turn, contributes to the well known DIS anomalous
dimension $B$~\cite{Sterman:1987aj,Catani:1989ne}. In the massive
case the virtuality of the final state is of order $(1-x)Q^2 + m^2$
and does not vanish in the soft limit which brings additional
$\ln(m)$ terms.
In this paper we will not elaborate on that point further, as all
purely collinear logarithmic terms can be read off from the
exponentiated expressions of the form factor given in
Eq.~(\ref{eq:massFFexp}). This picture is consistent with
fixed-order calculations of the perturbative fragmentation
function\footnote{We would like to thank S.~Catani for an
interesting discussion on this point.}. Indeed, one can easily
verify that the logarithmic contributions in the one-loop form
factor ${\cal F}_1$ for $\ep^0$ and $\ep$ coincide with the pure
virtual contributions to the one-loop fragmentation function $D_1$,
see for instance Eq.~(45) of Ref.~\cite{Melnikov:2004bm}.
Unfortunately, the two-loop virtual contributions cannot be
extracted from Ref.~\cite{Melnikov:2004bm}. In the next Section we
will elaborate on this relation.

Before completing this Section on the massive form factor, we would
like to address the question of its massless limit.
Reconsidering Eq.~(\ref{eq:massive-resummedff-explicit-form}),
i.e. the exponent of Eq.~(\ref{eq:massFFexp}),
it is obvious that the limit $m\to 0$ is smooth.
The contribution from the infrared function $K$ vanishes for $m\to 0$,
while the contribution from the $G$ function is similar to the one
in the well known massless case~\cite{Magnea:1990zb,Magnea:2000ss}.
The counterterm $K$ in Refs.~\cite{Magnea:1990zb,Magnea:2000ss}
can be expressed as
\begin{eqnarray}
\label{eq:K-couter-term}
K(0,\as,\epsilon) \, = \,
\int\limits_{0}^1\, {d x \over x}\  A({\bar a}(x \mu^2,\ep))
 \quad \, ,
\end{eqnarray}
which is consistent with Eqs.~(\ref{eq:massive-resummedff-explicit-form}) and (\ref{eq:massFFexp})
after changing the integration boundaries according to Refs.~\cite{Magnea:1990zb,Magnea:2000ss}.
Moreover, the function $G$ in the massive case should be related to the $G$ function
of the massless case up to possibly constant difference.
As we will demonstrate to two loops in the next Section,
the two functions in fact do coincide to all known orders in $\ep$,
i.e. the massless limit of the massive result
Eq.~(\ref{eq:massFFexp}) requires setting both $m=0$ and $C=1$.

Therefore, the resummed quark form factor reads in the massless case
\begin{eqnarray}
\label{eq:resummedff}
{\lefteqn{
\ln {\cal F}\!\left({Q^2 \over \mu^2},0,\as,\ep\right)
\, = \,
}}\\
&&
- {1 \over 2} \int\limits_0^{Q^2/\mu^2}\, {d \xi \over \xi}
\Bigg\{
  B(\bar a(\xi \mu^2,\ep)) +
  h(\bar a(\xi \mu^2,\ep)) +
  \int\limits_0^\xi\, {d \lambda \over \lambda}\
  A({\bar a}(\lambda \mu^2,\ep))
\Bigg\}
\quad \, ,
\nonumber
\end{eqnarray}
with the boundary condition ${\cal F}(0,0,\as,\ep) = 1$. Now $\ln
{\cal F}$ develops double poles in $\ep$ per power of $\ar$ from
the $\lambda$- and the $\xi$-integration over the anomalous
dimension $A$. In Eq.~(\ref{eq:resummedff}) we have identified the
initial condition $G\left({\bar a} \right)$
of Eq.~(\ref{eq:Gsol-massless}) with the sum of two functions $B + h$.
The physical interpretation of the new functions $B$ and $h$,
which also have expansions in the $d$-dimensional coupling, follows
nicely from the previous considerations of inclusive DIS scattering.
Following Ref.~\cite{Catani:1989ne} one can identify the function
$B$ with the coefficient governing the evolution of those large
logarithms $\ln (N)$ in inclusive DIS scattering associated with the
final state jet function. This has recently also been pointed out in
Ref.~\cite{Ravindran:2006cg}. The function $B$ is known to
three-loop accuracy \cite{Moch:2005ba} from explicit DIS
calculations~\cite{Moch:2002sn,Moch:2004pa,Vogt:2004mw,Vermaseren:2005qc}.
The new contribution $h$ on the other hand can be thought of as the
massless limit of the function $H$ in Eq.~(\ref{eq:res-FF}). To
determine $h$ in a perturbative expansion we match the above
exponent to the known three-loop result for the massless form
factor~\cite{Moch:2005id} and we have checked that the $\as^4$
prediction based on Eq.~(\ref{eq:resummedff}) agrees with previous
results in the literature~\cite{Moch:2005id}. It is interesting to
note that unlike the standard expression for the massless form
factor as given e.g. in
Refs.~\cite{Magnea:1990zb,Magnea:2000ss,Moch:2005id}, the result we
propose in this paper is comprised of well defined integrals.
Moreover, we can directly interpret the respective parts as the
three-loop contributions of the form factor to the DIS coefficient
functions and the splitting functions respectively.

Finally let us briefly comment again on the various flavor classes,
since the previous discussion was entirely focused on the
flavor classes $ll$ and $hl$. Beyond one loop
we have for instance the contributions to ${\cal F}$ from the class $hh$.
These contributions are finite after performing the ultraviolet
renormalization, but they still do contain Sudakov logarithms of
the type $\ln^k(Q^2/m^2)$. In fact, up to two loops all remaining
large logarithms $\ln^k(Q^2/m^2)$ in the heavy quark form factor
not accounted for by Eq.~(\ref{eq:massFFexp}) are entirely related
to the self-energy contributions of a heavy quark, i.e. the diagram
denoted with $hh$ in Fig.~\ref{fig:flavorclass}. It is well
known~\cite{Kuhn:2001hz,Feucht:2003yx} that these contributions
obey Sudakov exponentiation similar to Eq.~(\ref{eq:massFFexp}),
although with different integration boundaries and evaluated at the
matching scale $\mu^2 = m^2$. Thus, we can introduce the function
${\cal F}_{hh}$ which exponentiates the logarithms in the
flavor class $hh$,
\begin{eqnarray}
\label{eq:massFFexp-nf}
{\lefteqn{
\ln {\cal F}_{hh}(Q^2,m^2,{\bar a}(\mu^2,\ep),\ep) \, = \,
}}
\\&&
\left.
- {1 \over 2} \int\limits_{m^2/\mu^2}^{Q^2/\mu^2}\, {d \xi \over \xi}
\Bigg\{
  {G^\prime}({\bar a}(\xi \mu^2,\ep)) +
  K({\bar a}(\xi \mu^2 m^2/Q^2,\ep)) +
  \int\limits_{m^2/\mu^2}^\xi\, {d \lambda \over \lambda}\
  A({\bar a}(\lambda \mu^2,\ep))
\Bigg\}
\right|_{\nf=\nh}
\, . \nonumber
\end{eqnarray}
Eq.~(\ref{eq:massFFexp-nf}) is to be evaluated at the scale $\mu^2
= m^2$ and to be restricted to the purely fermionic contributions
with the heavy quark pair coupling to the external boson and one
additional virtual heavy quark line. Eq.~(\ref{eq:massFFexp-nf})
contains the same functions $A$ and $K$ as
Eq.~(\ref{eq:massFFexp}), but a different function $G^\prime$. As a
matter of fact, its structure follows directly from integrating
Eqs.~(\ref{eq:ffdeq}) and (\ref{eq:Gsol-massless}) under the
condition that the infrared region is cut off at the scale $m^2$.
See e.g. Refs.~\cite{Kuhn:2001hz,Feucht:2003yx} for details on the
finite-order expansion of exponentials like
Eq.~(\ref{eq:massFFexp-nf}).
We address this issue in future work.

%
% -----------------------------------------------------------------------------
%
\section{Fixed-order expansions and resummation coefficients}
\label{sec:rescoeff}

In this Section we will give the finite-order expansions for the various
quantities, in particular for the factor $Z^{(m\vert0)}$
of Eq.~(\ref{eq:Z}) and the form factor ${\cal F}$ of Eq.~(\ref{eq:massFFexp}).
All formulae in this Section use the $\MSbar$-scheme for the coupling
constant, while the heavy mass is always taken to be the pole mass
(on-shell scheme).
In addition, as in the previous Sections, we limit ourselves to the
contributions in the flavor classes $ll$ and $hl$.
Throughout this Section $\nf$ denotes the number of massless flavors.

\subsection{The factor $Z^{(m\vert0)}$}
Let us start with the perturbative QCD expansion of
Eq.~(\ref{eq:Z}) for the factor $Z^{(m\vert0)}$ which we defined as
the ratio of the massive and the massless form factors for a given
parton and which we write as an expansion in terms of the
renormalized coupling $\ar(\mu^2) \equiv \alpha_s(\mu^2) / (4\pi)$:
\begin{equation}
\label{eq:Z-exp} Z^{(m\vert0)}_{[i]}\left({m^2 \over \mu^2},\as,\ep
\right) \, = \, 1 + \sum\limits_{j=1}^{\infty} \left(\ar
\right)^j\, Z^{(j)}_{[i]} \, .
\end{equation}

The quark form factor in massless QCD is known to three
loops~\cite{Moch:2005id,Moch:2005tm}, while the form factor of a
massive quark is known for arbitrary values of the quark mass
through two loops~\cite{Bernreuther:2004ih}. The expansion
coefficients (\ref{eq:Z-exp}) in the case of a (heavy) quark $q$
read
\begin{eqnarray}
\label{eq:Zfactor-1loop}
  Z^{(1)}_{[q]} & = &
%%START
%%L %%texZQ1 =
\cf \* \biggl\{
  {2\over \ep^2}
+ {2\*L_{\mu} + 1 \over \ep}
+ L_{\mu}^2+L_{\mu}
+ 4
+ \z2
+ \ep \* \biggl(
  {L_{\mu}^3\over 3}
+ {L_{\mu}^2\over 2}
+ (4+\z2)\*L_{\mu}
+ 8
+ {\z2 \over 2}
- {2\over 3}\*\z3
  \biggr)
\\
& &\mbox{}
+ \ep^2 \* \biggl(
  {L_{\mu}^4 \over 12}
+ {L_{\mu}^3 \over 6}
+ \biggl(
  2
  + {\z2 \over 2}
  \biggr) \* L_{\mu}^2
+ \biggl(
  8
  + {\z2 \over 2}
- {2\over 3}\*\z3
  \biggr) \* L_{\mu}
+ 16
+ 2 \* \z2
- {\z3\over 3}
+ {9\over 20} \* \z2^2
  \biggr)
\biggr\}
%%;
%%STOP
\nonumber\\
& &\mbox{}
+ \calO(\ep^3)
\, ,
\nonumber\\
\label{eq:Zfactor-2loop}
  Z^{(2)}_{[q]} & = &
%%START
%%L %%texZQ2 =
  \cf^2\*{2\over \ep^4}
+ {1\over \ep^3}\*\biggl\{
  \cf^2\*(4\*L_{\mu}+2)
- {11\over 2}\*\cf\*\ca
+ \nf\*\cf\biggr\}
\\
& &\mbox{}
+ {1\over\ep^2}\*\biggl\{
    \cf^2\*\biggl(4\*L_{\mu}^2+4\*L_{\mu}+{17\over 2}+2\*\z2\biggr)
  + \cf\*\ca\*\biggl(-{11\over 3}\*L_{\mu} + {17\over 9}-\z2\biggr)
  + \nf\*\cf\*\biggl({2\over 3}\*L_{\mu}-{2\over 9}\biggr)
\biggr\}
\nonumber\\
& &\mbox{}
+{1\over\ep}\*\biggl\{
  \cf^2\*\biggl({8\over 3}\*L_{\mu}^3+4\*L_{\mu}^2+(17+4\*\z2)\*L_{\mu}
    +{83\over 4}-4\*\z2+{32\over 3}\*\z3\biggr)
\nonumber\\
& &\mbox{}
+ \cf\*\ca\*\biggl(\biggl({67\over 9}-2\*\z2\biggr)\*L_{\mu}
  +{373\over 108}+{15\over 2}\*\z2-15\*\z3\biggr)
+ \nf\*\cf\*\biggl(-{10\over 9}\*L_{\mu}-{5\over 54}-\z2\biggr)
\biggr\}
\nonumber\\
& &\mbox{}
+ \cf^2\*\biggl( {4\over 3}\*L_{\mu}^4+{8\over 3}\*L_{\mu}^3+(17+4\*\z2)\*L_{\mu}^2
    +\biggl({83\over 2}-8\*\z2+{64\over 3}\*\z3\biggr)\*L_{\mu}
    +{561 \over 8}
    +{61\over 2}\*\z2
    -{22\over 3}\*\z3
\nonumber\\
& &\mbox{}
    - 48\*\ln2\*\z2
    -{77\over 5}\*\z2^2
    \biggr)
+ \cf\*\ca\*\biggl(
     {11\over 9}\*L_{\mu}^3+\biggl({167\over 18}-2\*\z2\biggr)\*L_{\mu}^2
    +\biggl({1165\over 54}+{56\over 3}\*\z2-30\*\z3\biggr)\*L_{\mu}
\nonumber\\
& &\mbox{}
    +{12877\over 648}
    +{323\over 18}\*\z2
    +{89\over 9}\*\z3
    +24\*\ln2\*\z2
    -{47\over 5}\*\z2^2
    \biggr)
\nonumber\\
& &\mbox{}
+ \nf\*\cf\*\biggl(-{2\over 9}\*L_{\mu}^3-{13\over 9}\*L_{\mu}^2
    +\biggl(-{77\over 27}-{8\over 3}\*\z2\biggr)\*L_{\mu}
    -{1541\over 324}
    -{37\over 9}\*\z2
    -{26\over 9}\*\z3
    \biggr)
%%;
%%STOP
+ \calO(\ep)
\nonumber
\, ,
\end{eqnarray}
where two-loop contributions arising from
virtual heavy flavor lines are omitted (see Fig.~\ref{fig:flavorclass}) and
\begin{eqnarray}
\label{eq:Lmu-def}
L_{\mu} = \ln\left( {\mu^2\over m^2}\right)\, .
\end{eqnarray}

In presence of heavy flavors the form factor of a massless quark
gets mass-dependent contributions at two loops from the diagram
${\it lh}$ in Fig.~\ref{fig:flavorclass}. We will not consider such
two-loop contributions in this paper. Unlike the massless quark,
however, the gluon form factor receives mass-dependent corrections
starting from one loop. These have their origin in the one-loop heavy flavor
insertion in the tree-level gluon form factor. It is clear that for
the gluon form factor the classification of
Fig.~\ref{fig:flavorclass} has to be suitably adapted by counting
the number of (internal) heavy lines. It is easy to work out the
one-loop result for the gluon $Z^{(m\vert0)}$-factor in
Eq.~(\ref{eq:Z-exp}) (see the Appendix for the all-orders in $\ep$
result) and it reads:
\begin{eqnarray}
\label{eq:Zfactor-gluon}
  Z^{(1)}_{[g]} & = &
%%START
%%L %%texZg1 =
n_h \* \biggl\{
  -{2 \over 3 \* \ep} -{2 \over 3}\* L_{\mu} + \ep \*
\biggl(
  -{1\over 3} \* L_{\mu}^2 - {\z2 \over 3}
  \biggr)
 + \ep^2 \* \biggl(
  - {1 \over 9} \* L_{\mu}^3 - {\z2 \over 3} \* L_{\mu}
+ {2\*\z3\over 9}
  \biggr)
\biggr\}
%%;
%%STOP
+ \calO(\ep^3) \, .
\end{eqnarray}

In addition, the following comments on
Eqs.~(\ref{eq:Zfactor-1loop}), (\ref{eq:Zfactor-2loop}) above
are in order. First of all,
we have to supply the $\ep$-expansion of the massive form factor
including terms of order $\ep^2$ at one loop, because of the
singularity structure with $1/\ep^2$-poles in massless one-loop
form factor. Since the ${\cal O}(\ep^2)$ term of the one-loop
massive form factor is not available in the literature we have
calculated it following the setup of
Ref.~\cite{Bernreuther:2004ih}. Details are given in
Appendix~\ref{sec:appA}. One can easily verify that this term
produces a finite contribution at two loops for an amplitude with
$\nh$ external massive quarks, $\nl$ external massless quarks and
$\ng$ external gluons, which would be proportional to $\nl \cf^2 +
\ng \cf\ca$ times the Born term.

Secondly, there is one important detail about the scheme for
definition of coupling constant and masses. We assume the pole-mass
definition for the heavy quark mass $m$ as well as the standard
$\MSbar$ coupling defined in Eq.~(\ref{eq:alpha-s-renorm}). Note
that this definition for the coupling differs from
the one used in e.g. in Ref.~\cite{Bernreuther:2004ih} (and other
references on higher order corrections for massive processes) where
the coupling renormalization includes also the factor
$\Gamma(1+\ep)\exp(\ep\gamma_{\rm E})$.
For consistency with the massless calculations,
we have performed a finite renormalization of the
result in Ref.~\cite{Bernreuther:2004ih}. The necessary relation is
given by
\begin{eqnarray}
\label{eq:MSbar-recoupling}
\ar\biggr|_{\small \mbox{Ref.~\cite{Bernreuther:2004ih}}} & = &
\ar \left\{
  1 + \ar {1 \over \ep}
  \left(
  \beta_0 - {2 \over 3}
  \right)
  \left(
    {\Gamma(1+\ep) \over \exp(-\ep \gamma_{\rm E})} - 1
  \right)
  + \calO(\ar^2)
\right\}
\, ,
\end{eqnarray}
where we put the factor
$(4\pi)^\ep \exp(-\ep \gamma_{\rm E})=1$ for simplicity and
$\beta_0 = 11/3\ca - 2/3 \nf$.
It is easy to see that through two loops this amounts to the
following finite correction
(see Eq.~(\ref{eq:massFFpert}) below for definitions of ${\cal F}_i$)
to the results presented in Ref.~\cite{Bernreuther:2004ih}
\begin{eqnarray}
{\cal F}_2\biggr|_{\MSbar} & = & {\cal F}_2\biggr|_{\small
\mbox{Ref.~\cite{Bernreuther:2004ih}}}
+ \ar^2 {\beta_0 \over \ep}
\biggl( {\z2\over 2}\ep^2 + \calO(\ep^3)\biggr) \ {\cal
F}_1\biggr|_{\small
\mbox{Ref.~\cite{Bernreuther:2004ih}}}
\, .
\end{eqnarray}

Finally we would like to elaborate on the relation
between the factor $Z^{(m\vert0)}_{[q]}$ and the heavy quark
perturbative fragmentation function we discussed in
Section~\ref{sec:qcdffactor} preceding Eq.~(\ref{eq:resummedff}).
At one loop, the virtual contribution to the fragmentation function
was explicitly calculated in Ref.~\cite{Melnikov:2004bm} to all orders in $\ep$.
We present this result in Appendix~\ref{sec:appA} in a particular form prior
to collinear factorization and one can easily verify
by a direct comparison that its expansion
through ${\cal O}(\ep^2)$ coincides with the factor $Z^{(1)}_{[q]}$.
Moreover, Ref.~\cite{Melnikov:2004bm} also contains the purely virtual
fermionic contributions (i.e. proportional to the number of light flavors)
at two loops.
In terms of the usual renormalized coupling we have found the former to be in
agreement with the terms proportional to $\nf$ in the function $Z^{(2)}_{[q]}$
to all powers in $\ep$ appearing in Eq.~(\ref{eq:Zfactor-2loop}).
This observation indicates that the factor $Z^{(m\vert0)}_{[q]}$
of Eq.~(\ref{eq:Z}) indeed coincides with the virtual corrections to the
collinearly unfactorized perturbative fragmentation function and
one may actually view the complete agreement between all known terms of the two
functions as a check on the derivation of $Z^{(m\vert0)}_{[q]}$.

Although the latter object is not known to the level we have presented here
for the function $Z^{(m\vert0)}_{[q]}$ the apparent coincidence allows for
an interesting alternative interpretation of that function
by relating it to the field renormalization constant of a heavy quark
in light cone gauge $n\cdot A=0$.
Indeed, in the approach of Ref.\cite{Melnikov:2004bm} to calculate
the fragmentation function, the purely virtual corrections are nothing but
insertions of self-energy type in external on-shell legs in this particular gauge.
Clearly, it will be very interesting to further develop this line of reasoning.

\subsection{The form factor ${\cal F}$}
Next, we want to perform the finite-order expansion
and matching of Eq.~(\ref{eq:massFFexp})
for the heavy quark form factor ${\cal F}$.
Subsequently, with all functions $G$, $K$ and $F$ determined
we will then be using Eq.~(\ref{eq:massFFexp})
for predictions of perturbative results at higher orders and
derive explicit results at three loops.
To that end we perform the integrations in Eq.~(\ref{eq:massive-resummedff}) after
inserting the perturbative expansions of all quantities and simply
evaluated resulting integrals.
Details on this procedure may be found in Refs.~\cite{Magnea:2000ss,Moch:2005id}.

For the (ultraviolet) renormalized massive form factor
with space-like virtuality $q^2 = - Q^2 <0$ and
in terms of the renormalized coupling $\as(\mu^2)$
we have,
\begin{equation}
  \label{eq:massFFpert}
{\cal F}\left({Q^2 \over \mu^2},{m^2 \over \mu^2},\ar(\mu^2),\ep \right)
\, = \,
1 + \sum\limits_{i=1}^{\infty}
\left(\ar \right)^i\, {\cal F}_i
\, .
\end{equation}
With the convention of Eq.~(\ref{eq:abexp}) for the expansion of
$A$, $G$, $K$ and $C$ and setting the scale to $\mu^2 = m^2$,
we find
\begin{eqnarray}
  \label{eq:massF1}
{\cal F}_1 &=&
%%START
%%L %%texF1 =
         {1 \over \ep} \* \biggl\{
            {1 \over 2} \* A_1 \* L
          + {1 \over 2} \* (
            G_1
          + K_1
          )
         \biggr\}
       - {1 \over 4} \* A_1 \* L^2
       - {1 \over 2} \* G_1 \* L
       + C_1
       + \ep \* \biggl\{
            {1 \over 12} \* A_1 \* L^3
          + {1 \over 4} \* G_1 \* L^2
       \biggr\}
\\
& &\mbox{}
       - \ep^2 \*  \biggl\{
            {1 \over 48} \* A_1 \* L^4
          + {1 \over 12} \* G_1 \* L^3
       \biggr\}
%%;
%%STOP
       + \calO(\ep^3)
\nonumber\, ,
\\[1ex]
\label{eq:massF2}
{\cal F}_2 &=&
%%START
%%L %%texF2 =
         {1 \over \ep^2} \* \biggl\{
         {1 \over 8} \* A_1^2 \* L^2
       + {1 \over 4} \*  A_1 \* (
            G_1
          + K_1
          - \beta_0
          ) \* L
       + {1 \over 8}  \* (
            G_1
          + K_1
          ) \* (
            G_1
          + K_1
          - 2 \* \beta_0
          )
          \biggr\}
\\
& &\mbox{}
       + {1 \over \ep} \* \biggl\{
       - {1 \over 8} \* A_1^2 \* L^3
       - {1 \over 8} \* A_1  \*  (
            3 \* G_1
          + K_1
          ) \* L^2
       + {1 \over 4} \* (
            A_2
          - G_1^2
          - K_1 \* G_1
          + 2 \* A_1 \* C_1
          ) \* L
       + {1 \over 4} \* (
            G_2
          + K_2
          )
\nonumber\\
& &\mbox{}
       + {1 \over 2} \* C_1 \* (
            G_1
          + K_1
          )
          \biggr\}
       + {7 \over 96} \* A_1^2 \* L^4
       + {1 \over 24} \* A_1 \* (
            7 \* G_1
          + K_1
          + 2 \* \beta_0
          ) \* L^3
       + {1 \over 8} \*
            G_1 \* (
            2 \* G_1
          + K_1
          + 2 \* \beta_0
          ) \* L^2
\nonumber\\
& &\mbox{}
       - {1 \over 4} \* (
            A_2
          + A_1 \* C_1
          ) \* L^2
       - {1 \over 2} \* (
            G_2
          + G_1 \* C_1
          ) \* L
       + C_2
       + \ep \* \biggl\{
       - {1 \over 32} \* A_1^2 \* L^5
       - {1 \over 96} \* A_1 \* (
           15 \* G_1
          + K_1
          + 6 \* \beta_0
          ) \* L^4
\nonumber\\
& &\mbox{}
       - {1 \over 24} \* G_1 \* (
            4 \* G_1
          + K_1
          + 6\* \beta_0
          ) \* L^3
       + {1 \over 12} \* (
            2 \* A_2
          + A_1 \* C_1
          ) \* L^3
       + {1 \over 4} \* (
            2 \* G_2
          + G_1 \* C_1
          ) \* L^2
       \biggr\}
%%;
%%STOP
       + \calO(\ep^2)
\nonumber\, ,
\\[1ex]
\label{eq:massF3}
{\cal F}_3 &=&
%%START
%%L %%texF3 =
         {1 \over \ep^3} \* \biggl\{
         {1 \over 48} \* A_1^3 \* L^3
       + {1 \over 16} \* A_1^2 \* (
            G_1
          + K_1
          - 2 \* \beta_0
          ) \* L^2
       + {1 \over 16} \* A_1 \* (
            G_1
          + K_1
          ) \* (
            G_1
          + K_1
          - 4 \* \beta_0
          ) \* L
\\
& &\mbox{}
       + {1 \over 6} \* A_1 \* \beta_0^2 \* L
       +  {1 \over 48} \* (
            G_1
          + K_1
          ) \* (
            G_1
          + K_1
          - 2 \* \beta_0
          ) \* (
            G_1
          + K_1
          - 4 \* \beta_0
          )
          \biggr\}
       + {1 \over \ep^2} \* \biggl\{
       - {1 \over 32} \* A_1^3 \* L^4
\nonumber\\
& &\mbox{}
       - {1 \over 16} \* A_1^2 \* (
            2 \* G_1
          + K_1
          - \beta_0
          ) \* L^3
       + {1 \over 8} \* A_1 \* (
            A_2
          + A_1 \* C_1
          ) \* L^2
       + {1 \over 16} \* A_1 \* \beta_0 \* (
            3 \* G_1
          + K_1
          ) \* L^2
\nonumber\\
& &\mbox{}
       - {1 \over 32} \* A_1 \* (
            G_1
          + K_1
          ) \* (
            5 \* G_1
          + K_1
          ) \* L^2
       + {1 \over 24} \* A_2 \* (
            3 \* G_1
          + 3 \* K_1
          - 4 \* \beta_0
          ) \* L
       + {1 \over 24} \* A_1 \* (
            3 \* G_2
          + 3 \* K_2
          - 4 \* \beta_1
          ) \* L
\nonumber\\
& &\mbox{}
       - {1 \over 16} \* G_1 \* (
            G_1
          + K_1
          ) \* (
            G_1
          + K_1
          - 2 \* \beta_0
          ) \* L
          + {1 \over 4} \* A_1 \* C_1 \* (
            G_1
          + K_1
          - \beta_0
          ) \* L
       + {1 \over 8} \* C_1 \* (
            G_1
          + K_1
          )^2
\nonumber\\
& &\mbox{}
       + {1 \over 24} \* (
            G_1
          + K_1
          ) \* (
            3 \* G_2
          + 3 \* K_2
          - 6 \* \beta_0 \* C_1
          - 4 \* \beta_1
          )
      - {1 \over 6} \* \beta_0 \* (
            G_2
          + K_2
          )
          \biggr\}
       + {1 \over \ep} \* \biggl\{
         {5 \over 192} \* A_1^3 \* L^5
\nonumber\\
& &\mbox{}
       + {1 \over 192} \* A_1^2 \* (
            25 \* G_1
          + 7 \* K_1
          + 4 \* \beta_0
          ) \* L^4
       + {1 \over 96} \* A_1 \* (
            19 \* G_1^2
          + K_1^2
          + 14 \* K_1 \* G_1
          ) \* L^3
\nonumber\\
& &\mbox{}
       + {1 \over 48} \* A_1 \* \beta_0 \* (
            4 \* G_1
          + K_1
          ) \* L^3
       - {1 \over 16} \* A_1 \* (
            3 \* A_2
          + 2 \* A_1 \* C_1
          ) \* L^3
       + {1 \over 32} \* G_1 \* (
            G_1
          + K_1
          ) \* (
            3 \* G_1
          + K_1
          + 2 \* \beta_0
          ) \* L^2
\nonumber\\
& &\mbox{}
          - {1 \over 8} \* A_2 \* (
            2 \* G_1
          + K_1
          ) \* L^2
          - {1 \over 16} \* A_1 \* (
            5 \* G_2
          + K_2
          + 6 \* G_1 \* C_1
          + 2 \* K_1 \* C_1
          ) \* L^2
       + {1 \over 36} \* A_1 \* (
            32 \* \ca
          - 135 \* \cf
          ) \* L
\nonumber\\
& &\mbox{}
       + {1 \over 12} \* (
            2\* A_3
          + 3 \* A_2 \* C_1
          + 6 \* A_1 \* C_2
          ) \* L
       + {1 \over 8} \* (
          - 3 \* G_1 \* G_2
          - 2 \* K_1 \* G_2
          - K_2 \* G_1
          ) \* L
       - {1 \over 4} \* G_1 \* C_1 \* (
            G_1
          + K_1
          ) \* L
\nonumber\\
& &\mbox{}
          + {1\over 6} \* (
            G_3
          + K_3
          )
          + {1 \over 4} \* C_1 \* (
            G_2
          + K_2
          )
        + {1 \over 36} \* (
            G_1
          + K_1
          ) \* (
            18 \* C_2
          + 32 \* \ca
          - 135 \* \cf
          )
          \biggr\}
\nonumber\\
& &\mbox{}
       - {1 \over 64} \* A_1^3 \* L^6
       - {1 \over 64} \* A_1^2 \* (
            6 \* G_1
          + K_1
          + 3 \* \beta_0
          ) \* L^5
       + {1 \over 96} \* A_1 \* (
            16 \* A_2
          + 7 \* A_1 \* C_1
          ) \* L^4
       - {1 \over 384} \* A_1 \* (
            65 \* G_1^2
\nonumber\\
& &\mbox{}
          + 30 \* K_1 \* G_1
          + K_1^2
          + 90 \* \beta_0 \* G_1
          + 10 \* \beta_0 \* K_1
          + 16 \* \beta_0^2
          ) \* L^4
       + {1 \over 48} \* A_2 \* (
            13 \* G_1
          + 4 \* K_1
          + 8 \* \beta_0
          ) \* L^3
\nonumber\\
& &\mbox{}
       + {1 \over 48} \* A_1 \* (
            19 \* G_2
          + K_2
          + 4 \* \beta_1
          ) \* L^3
       + {1 \over 24} \* A_1 \* C_1 \* (
            7 \* G_1
          + K_1
          + 2 \* \beta_0
          ) \* L^3
       - {1 \over 96} \* G_1 \* (
            9 \* G_1^2
          + K_1^2
\nonumber\\
& &\mbox{}
          + 8 \* K_1 \* G_1
          + 22 \* \beta_0 \* G_1
          + 10 \* \beta_0 \* K_1
          + 16 \* \beta_0^2
          ) \* L^3
          - {1 \over 4} \* (
            A_3
          + A_2 \* C_1
          ) \* L^2
          + {1 \over 16} \* G_1 \* (
            K_2
          + 4 \* \beta_1
            ) \* L^2
\nonumber\\
& &\mbox{}
          - {1 \over 72} \* A_1 \* (
             18 \* C_2
          + 32 \* \ca
          - 135 \* \cf
            ) \* L^2
          + {1 \over 16} \* G_2 \* (
            9 \* G_1
          + 4 \* K_1
          + 8 \* \beta_0
          ) \* L^2
\nonumber\\
& &\mbox{}
          + {1 \over 8} \* G_1 \* C_1 \* (
            2 \* G_1
          + K_1
          + 2 \* \beta_0
          ) \* L^2
       - {1 \over 2} \* (
           \tdG_3
         + G_2 \* C_1
         + G_1 \* C_2
         ) \* L
       + \tdC_3
%%;
%%STOP
       + \calO(\ep)
\nonumber\, ,
\end{eqnarray}
where again contributions arising from virtual heavy flavor lines are omitted
(class $hh$) and
\begin{eqnarray}
\label{eq:L-def}
L = \ln\left( {Q^2\over m^2}\right)\, .
\end{eqnarray}
In the quantities $\tdG_3$ and $\tdC_3$ in Eq.~(\ref{eq:massF3})
we have also absorbed all constant contributions from the decoupling relation~(\ref{eq:adecd})
at order $\ar^3$.
All these terms are independent of $L_{\mu, \ep}$ and can potentially include contributions
of order $\ep$ at $\ar^3$ which we did not write out explicitly in Eq.~(\ref{eq:adecd}).
Results for ${\cal F}_i$ at a general scale $\mu^2 \neq m^2$ can be derived
from Eqs.~(\ref{eq:massF1})--(\ref{eq:massF3}) by standard methods\footnote{
We take the opportunity to point out a typographical mistake in Eq.~(62) of Ref.~\cite{Bernreuther:2004ih}.
The following term
$\displaystyle \ca \cf {11 \over 6}
\biggl({3 \over 2} - {1 \over 2(1-x)} - {1 \over (1+x)} \biggr) H(0,x) $
should actually read
$\displaystyle \ca \cf {11 \over 6}
\biggl({3 \over 2} - {2 \over (1-x)} - {1 \over (1+x)} \biggr) H(0,x) $.
}.
They will be presented elsewhere.

Explicit results for ${\cal F}_i$ in Eqs.~(\ref{eq:massF1})--(\ref{eq:massF3})
can be obtained with the help of the known coefficients of
the cusp anomalous dimension $A(\ar)$
due to Refs.~\cite{Moch:2004pa,Vogt:2004mw,Kodaira:1982nh},
\begin{eqnarray}
\label{eq:a1}
  A_1 & = &
%%START
%%L %%texA1 =
            4 \* \cf
%%;
%%STOP
\, ,
\\
\label{eq:a2}
  A_2 & = &
%%START
%%L %%texA2 =
         \cf \* \ca  \*  \biggl(
            {268 \over 9}
          - 8 \* \z2
          \biggr)
       + \nf \* \cf  \*  \biggl(
          - {40 \over 9}
          \biggr)
%%;
%%STOP
\, ,
\\
\label{eq:a3}
  A_3 & = &
%%START
%%L %%texA3 =
         \cf \* \ca^2  \*  \biggl(
            {490 \over 3}
          - {1072 \over 9} \* \z2
          + {88 \over 3} \* \z3
          + {176 \over 5} \* \z2^2
          \biggr)
\\
& &\mbox{}
       + \nf \* \cf \* \ca  \*  \biggl(
          - {836 \over 27}
          + {160 \over 9} \* \z2
          - {112 \over 3} \* \z3
          \biggr)
       + \nf \* \cf^2  \*  \biggl(
          - {110 \over 3}
          + 32 \* \z3
          \biggr)
       + \nf^2 \* \cf  \*  \biggl(
          - {16 \over 27}
          \biggr)
%%;
%%STOP
\, .
\nonumber
\end{eqnarray}
The respective coefficients for $G(\ar)$ and $K(\ar)$ read,
\begin{eqnarray}
\label{eq:g1}
  G_1 & = &
%%START
%%L %%texG1 =
          - 6 \* \cf
       + \ep \* \cf \* (
          - 16
          + 2 \* \z2
          )
       + \ep^2 \* \cf \* \biggl(
          - 32
          + 3 \* \z2
          + {28 \over 3} \* \z3
          \biggr)
\\
& &\mbox{}
       + \ep^3 \* \cf \* \biggl(
          - 64
          + 8 \* \z2
          + 14 \* \z3
          + {47 \over 10} \* \z2^2
          \biggr)
%%;
%%STOP
+ \calO\left(\ep^4\right)
\, ,
\\
\label{eq:g2}
  G_2 & = &
%%START
%%L %%texG2 =
         \cf^2  \*  (
          - 3
          + 24 \* \z2
          - 48 \* \z3
          )
       + \cf \* \ca  \*  \biggl(
          - {2545 \over 27}
          - {44 \over 3} \* \z2
          + 52 \* \z3
          \biggr)
\\
& &\mbox{}
       + \nf \* \cf  \*  \biggl(
            {418 \over 27}
          + {8 \over 3} \* \z2
          \biggr)
%%;
%%STOP
+ \calO\left(\ep\right)
\nonumber \, ,
\\
\label{eq:k1}
  K_1 & = &
%%START
%%L %%texK1 =
           2 \* \cf
%%;
%%STOP
\, ,
\\
\label{eq:k2}
  K_2 & = &
%%START
%%L %%texK2 =
         \cf^2  \*  (
            3
          - 24 \* \z2
          + 48 \* \z3
          )
       + \cf \* \ca  \*  \biggl(
            {373 \over 27}
          + 30 \* \z2
          - 60 \* \z3
          \biggr)
       + \nf \* \cf  \*  \biggl(
          - {10 \over 27}
          - 4 \* \z2
          \biggr)
%%;
%%STOP
\, .
\end{eqnarray}
Here we have included higher orders of $\ep$ in the anomalous dimensions,
to ensure that all large logarithms in $m$ are actually generated
entirely by the integrations over $\xi$ and $\lambda$ in
Eq.~(\ref{eq:massFFexp}).
Although this is not a compelling choice it captures
all structures in the exponential, which are universally
related to parton dynamics.
This is contrary to ``minimal'' versions proposed e.g.
in Ref.~\cite{MertAybat:2006mz}.

For the coefficients of the matching function $C(\ar)$ we find,
\begin{eqnarray}
\label{eq:c1}
  C_1 & = &
%%START
%%L %%texC1 =
         \cf \* (4 + \z2)
       + \ep \* \cf \* \biggl(
            8
          + {1 \over 2} \* \z2
          - {2 \over 3} \* \z3
          \biggr)
       + \ep^2 \* \cf \* \biggl(
            16
          + 2 \* \z2
          - {1 \over 3} \* \z3
          + {9 \over 20} \* \z2^2
          \biggr)
%%;
%%STOP
+ \calO\left(\ep^3\right)
\, ,
\quad
\\
\label{eq:c2}
  C_2 & = &
%%START
%%L %%texC2 =
         \cf^2 \* \biggl(
            30
          + 55 \* \z2
          - 36 \* \z3
          - 48 \* \z2 \* \ln2
          - {251 \over 10} \* \z2^2
          \biggr)
\\
& &\mbox{}
       + \ca \* \cf \* \biggl(
          - {2387 \over 27}
          + {71 \over 36} \* \z2
          + {479 \over 9} \* \z3
          + 24 \* \z2 \* \ln2
          - {3 \over 5} \* \z2^2
          \biggr)
\nonumber\\
& &\mbox{}
       + \nf \* \cf \* \biggl(
            {356 \over 27}
          - {37 \over 18} \* \z2
          - {38 \over 9} \* \z3
          \biggr)
%%;
%%STOP
+ \calO\left(\ep\right)
\nonumber
\, .
\end{eqnarray}
Putting everything together,
including the terms of order $\ep^2$ at one loop (see Appendix~\ref{sec:appA})
we arrive at the following results,
\begin{eqnarray}
  \label{eq:massF1explicit}
{\cal F}_1 &=&
%%START
%%L %%texF1explicit =
         \cf \* \biggl\{
         {1 \over \ep} \* (
            2 \* L
          - 2
         )
          - L^2
          + 3 \* L
          - 4
          + 2 \* \z2
       + \ep \* \biggl(
            {1 \over 3} \* L^3
          - {3 \over 2} \* L^2
          + (
            8
          - \z2 ) \* L
          - 8
          + 2 \* \z2
          + 4 \* \z3
       \biggr)
\\
& &\mbox{}
       + \ep^2 \* \biggl(
          - {1 \over 12} \* L^4
          + {1 \over 2} \* L^3
          - \biggl(
            4
          - {1 \over 2} \* \z2 \biggr) \* L^2
          + \biggl(
            16
          - {3 \over 2} \* \z2
          - {14 \over 3} \* \z3 \biggr) \* L
\nonumber\\
& &\mbox{}
          - 16
          + 6 \* \z2
          + {20 \over 3} \* \z3
          + {14 \over 5} \* \z2^2
       \biggr)
       \biggr\}
%%;
%%STOP
       + \calO(\ep^3)
\nonumber\, ,
\\[1ex]
\label{eq:massF2explicit}
{\cal F}_2 &=&
%%START
%%L %%texF2explicit =
         \cf^2 \* \biggl\{
         {1 \over \ep^2} \* (
            2 \* L^2
          - 4 \* L
          + 2
          )
       + {1 \over \ep} \* (
          - 2 \* L^3
          + 8 \* L^2
          - (
            14
          - 4 \* \z2 ) \* L
          + 8
          - 4 \* \z2
          )
\\
& &\mbox{}
          + {7 \over 6} \* L^4
          - {20 \over 3} \* L^3
          + \biggl(
            {55 \over 2}
          - 4 \* \z2
            \biggr) \* L^2
          - \biggl(
            {85 \over 2}
          - 32 \* \z3
            \biggr) \* L
          + 46
          + 39 \* \z2
          - 44 \* \z3
          - 48 \* \z2 \* \ln2
\nonumber\\
& &\mbox{}
          - {118 \over 5} \* \z2^2
       + \ep \* \biggl(
          - {1 \over 2} \* L^5
          + {11 \over 3} \* L^4
          - \biggl(
            {137 \over 6}
          - {8 \over 3} \* \z2
            \biggr) \* L^3
          + \biggl(
            {153 \over 2}
          - {112 \over 3} \* \z3
            \biggr) \* L^2
          \biggr)
       \biggr\}
\nonumber\\
& &\mbox{}
       + \ca \* \cf \* \biggl\{
         {1 \over \ep^2} \* \biggl(
          - {11 \over 3} \* L
          + {11 \over 3}
          \biggr)
       + {1 \over \ep} \* \biggl(
          \biggl(
            {67 \over 9}
          - 2 \* \z2
          \biggr) \* L
          - {49 \over 9}
          + 2 \* \z2
          - 2 \* \z3
          \biggr)
          + {11 \over 9} \* L^3
\nonumber\\
& &\mbox{}
          - \biggl(
            {233 \over 18}
          - 2 \* \z2
            \biggr) \* L^2
          + \biggl(
            {2545 \over 54}
          + {22 \over 3} \* \z2
          - 26 \* \z3
            \biggr) \* L
          - {1595 \over 27}
          - {7 \over 9} \* \z2
          + {134 \over 3} \* \z3
          + 24 \* \z2 \* \ln2
\nonumber\\
& &\mbox{}
          - {3 \over 5} \* \z2^2
       + \ep \* \biggl(
          - {11 \over 12} \* L^4
          + \biggl(
            {565 \over 54}
          - {4 \over 3} \* \z2
            \biggr) \* L^3
          - \biggl(
            {3337 \over 54}
          + {11 \over 2} \* \z2
          - 26 \* \z3
            \biggr) \* L^2
          \biggr)
       \biggr\}
\nonumber\\
& &\mbox{}
       + \cf \* \nf \* \biggl\{
         {1 \over \ep^2} \* \biggl(
            {2 \over 3} \* L
          - {2 \over 3}
          \biggr)
       + {1 \over \ep} \* \biggl(
          - {10 \over 9} \* L
          + {10 \over 9}
          \biggr)
          - {2 \over 9} \* L^3
          + {19 \over 9} \* L^2
          - \biggl(
            {209 \over 27}
          + {4 \over 3} \* \z2
            \biggr) \* L
          + {212 \over 27}
\nonumber\\
& &\mbox{}
          - {14 \over 9} \* \z2
          - {8 \over 3} \* \z3
       + \ep \* \biggl(
            {1 \over 6} \* L^4
          - {47 \over 27} \* L^3
          + \biggl(
            {281 \over 27}
          + \z2
            \biggr) \* L^2
          \biggr)
      \biggr\}
%%;
%%STOP
       + \calO(L\, \ep)
       + \calO(\ep^2)
\nonumber\, ,
\\[1ex]
\label{eq:massF3explicit}
{\cal F}_3 &=&
%%START
%%L %%texF3explicit =
         \cf^3 \* \biggl\{
         {1 \over \ep^3} \* \biggl(
            {4 \over 3} \* L^3
          - 4 \* L^2
          + 4 \* L
          - {4 \over 3}
          \biggr)
       + {1 \over \ep^2} \* (
          - 2 \* L^4
          + 10 \* L^3
          - (
            22
          - 4 \* \z2 ) \* L^2
\\
& &\mbox{}
          + (
            22
          - 8 \* \z2 ) \* L
          - 8
          + 4 \* \z2
          )
       + {1 \over \ep} \* \biggl(
            {5 \over 3} \* L^5
          - {34 \over 3} \* L^4
          + \biggl(
            {137 \over 3}
          - 6 \* \z2
          \biggr) \* L^3
          - \biggl(
            89
          - 56 \* \z3
          \biggr) \* L^2
\nonumber\\
& &\mbox{}
          + \biggl(
            129
          + 88 \* \z2
          - 136 \* \z3
          - 96 \* \z2 \* \ln2
          - {236 \over 5} \* \z2^2
          \biggr) \* L
          \biggr)
          - L^6
          + {17 \over 2} \* L^5
          - \biggl(
            {148 \over 3}
          - {16 \over 3} \* \z2
          \biggr) \* L^4
\nonumber\\
& &\mbox{}
          + \biggl(
            {494 \over 3}
          + {17 \over 3} \* \z2
          - {268 \over 3} \* \z3
          \biggr) \* L^3
       \biggr\}
       + \cf^2 \* \ca \* \biggl\{
         {1 \over \ep^3} \* \biggl(
          - {22 \over 3} \* L^2
          + {44 \over 3} \* L
          - {22 \over 3}
          \biggr)
       + {1 \over \ep^2} \* \biggl(
            {11 \over 3} \* L^3
\nonumber\\
& &\mbox{}
          + \biggl(
            {2 \over 9}
          - 4 \* \z2
          \biggr) \* L^2
          - \biggl(
            {1 \over 9}
          - {2 \over 3} \* \z2
          + 4 \* \z3
          \biggr) \* L
          - {34 \over 9}
          + {10 \over 3} \* \z2
          + 4 \* \z3
          \biggr)
       + {1 \over \ep} \* \biggl(
            {11 \over 9} \* L^4
\nonumber\\
& &\mbox{}
          - \biggl(
            {523 \over 18}
          - 6 \* \z2
          \biggr) \* L^3
          + \biggl(
            {6107 \over 54}
          + {19 \over 3} \* \z2
          - 50 \* \z3
          \biggr) \* L^2
          - \biggl(
            {5396 \over 27}
          - {5 \over 3} \* \z2
          - {362 \over 3} \* \z3
          - 48 \* \z2 \* \ln2
\nonumber\\
& &\mbox{}
          + {26 \over 5} \* \z2^2
          \biggr) \* L
          \biggr)
          - {11 \over 4} \* L^5
          + \biggl(
            {4289 \over 108}
          - {16 \over 3} \* \z2
          \biggr) \* L^4
          - \biggl(
            {6260 \over 27}
          + {97 \over 18} \* \z2
          - {232 \over 3} \* \z3
          \biggr) \* L^3
       \biggr\}
\nonumber\\
& &\mbox{}
       + \cf \* \ca^2 \* \biggl\{
         {1 \over \ep^3} \* \biggl(
            {242 \over 27} \* L
          - {242 \over 27}
          \biggr)
       + {1 \over \ep^2} \* \biggl(
          - \biggl(
            {2086 \over 81}
          - {44 \over 9} \* \z2
          \biggr) \* L
          + {1690 \over 81}
          - {44 \over 9} \* \z2
          + {44 \over 9} \* \z3
          \biggr)
\nonumber\\
& &\mbox{}
       + {1 \over \ep} \* \biggl(
          \biggl(
            {245 \over 9}
          - {536 \over 27} \* \z2
          + {44 \over 9} \* \z3
          + {88 \over 15} \* \z2^2
          \biggr) \* L
          \biggr)
          - {121 \over 54} \* L^4
          + \biggl(
            {2869 \over 81}
          - {44 \over 9} \* \z2
          \biggr) \* L^3
       \biggr\}
\nonumber\\
& &\mbox{}
       + \cf^2 \* \nf \* \biggl\{
         {1 \over \ep^3} \* \biggl(
            {4 \over 3} \* L^2
          - {8 \over 3} \* L
          + {4 \over 3}
          \biggr)
       + {1 \over \ep^2} \* \biggl(
          - {2 \over 3} \* L^3
          + {4 \over 9} \* L^2
          + \biggl(
            {10 \over 9}
          + {4 \over 3} \* \z2
          \biggr) \* L
          - {8 \over 9}
          - {4 \over 3} \* \z2
          \biggr)
\nonumber\\
& &\mbox{}
       + {1 \over \ep} \* \biggl(
          - {2 \over 9} \* L^4
          + {41 \over 9} \* L^3
          - \biggl(
            {481 \over 27}
          + {10 \over 3} \* \z2
          \biggr) \* L^2
          + \biggl(
            {599 \over 27}
          - {2 \over 3} \* \z2
          + {8 \over 3} \* \z3
          \biggr) \* L
          \biggr)
          + {1 \over 2} \* L^5
          - {355 \over 54} \* L^4
\nonumber\\
& &\mbox{}
          + \biggl(
            {1016 \over 27}
          + {29 \over 9} \* \z2
            \biggr) \* L^3
       \biggr\}
       + \cf \* \ca \* \nf \* \biggl\{
         {1 \over \ep^3} \* \biggl(
          - {88 \over 27} \* L
          + {88 \over 27}
          \biggr)
       + {1 \over \ep^2} \* \biggl(
            \biggl(
            {668 \over 81}
          - {8 \over 9} \* \z2
          \biggr) \* L
\nonumber\\
& &\mbox{}
          - {596 \over 81}
          + {8 \over 9} \* \z2
          - {8 \over 9} \* \z3
          \biggr)
       + {1 \over \ep} \* \biggl(
          - \biggl(
            {418 \over 81}
          - {80 \over 27} \* \z2
          + {56 \over 9} \* \z3
          \biggr) \* L
          \biggr)
          + {22 \over 27} \* L^4
          - \biggl(
            {974 \over 81}
          - {8 \over 9} \* \z2
          \biggr) \* L^3
       \biggr\}
\nonumber\\
& &\mbox{}
       + \cf \* \nf^2 \* \biggl\{
         {1 \over \ep^3} \* \biggl(
            {8 \over 27} \* L
          - {8 \over 27}
          \biggr)
       + {1 \over \ep^2} \* \biggl(
          - {40 \over 81} \* L
          + {40 \over 81}
          \biggr)
       + {1 \over \ep} \* \biggl(
          - {8 \over 81} \* L
          \biggr)
       - {2 \over 27} \* L^4
       + {76 \over 81} \* L^3
       \biggr\}
\nonumber\\
& &\mbox{}
       + \cf^2 \* \biggl\{
         {1 \over \ep} \* \biggl(
          - 15 \* L
          \biggr)
       \biggr\}
       + \cf \* \ca \* \biggl\{
         {1 \over \ep} \* \biggl(
            {32 \over 9} \* L
          \biggr)
       \biggr\}
%%;
%%STOP
       + \calO(L^0\, \ep^{-1})
       + \calO(L^2\, \ep^0)
       + \calO(\ep)
\nonumber
\, .
\end{eqnarray}
Further improvements on the accuracy of the three-loop prediction
${\cal F}_3$ require an extension of the two-loop result ${\cal F}_2$
to order $\ep$. We will return to this issue in a future publication.

Let us close this Section with a few comments.
First of all, it is clear we can obtain a three-loop prediction,
i.e. the coefficient $Z^{(3)}_{[q]}$ in Eq.~(\ref{eq:Z-exp})
for the factor $Z^{(m\vert0)}_{[q]}$ from the exponentiated massive form factor
in Eq.~(\ref{eq:massFFexp}) with the help of Eq.~(\ref{eq:massF3explicit})
and the known three-loop results in the massless case~\cite{Moch:2005id,Moch:2005tm}.
Next, putting the discussion in a broader perspective, we note that
exponentiations similar to Eq.~(\ref{eq:massFFexp}) have also been studied for
electroweak interactions~\cite{Kuhn:2001hz,Jantzen:2005az}
in massive gauge theories, where large logarithms in the mass of the gauge boson appear.
There, the resummation has been used as a generating functional
for Sudakov logarithms at higher orders.
Of course, those details of the exponentials which depend on the infrared sector
of the theory are modified in comparison to Eq.~(\ref{eq:massFFexp}).
However, it is rather striking to observe that the coefficients for $G_1$ and $G_2$
from our determination in Eqs.~(\ref{eq:g1}) and (\ref{eq:g2})
agree precisely with the values for $\zeta^{(1)}$ , $\zeta^{(2)}$
in Ref.~\cite{Jantzen:2005az}
(up to an overall factor $1/2$ due to different normalizations).
In both cases, the relevant coefficients control the single logarithm $L$
at the respective order.

On top of this, it is even more striking, that the very same coefficients
$G_1$ and $G_2$ from Eqs.~(\ref{eq:g1}), (\ref{eq:g2}) for the form factor of
a massive quark also coincide with the corresponding results
in massless case~\cite{Moch:2005id,Moch:2005tm}
(up to an overall sign from different definitions).
This observation, which calls for an explanation,
suggests a universality of the function $G$ which
extends even to higher orders in $\ep$, see e.g. Eq.~(\ref{eq:g1}).
It also offers the chance for a conjecture about the coefficient of the single
logarithm $L$ for ${\cal F}_3$ in Eq.~(\ref{eq:massF3}) purely on the basis
of the corresponding massless result, provided, of course,
that all necessary terms to higher order in $\ep$ up to two loops are known.

The potential consequences of such a universal nature of the function $G$
would be rather interesting.
For instance, in the massless case there exist additional relations between the
functions ${\cal F}^{[q]}$ and ${\cal F}^{[g]}$, i.e. the form factors
for the vertices $\gamma^{\,\ast}\!qq$ and $\phi gg$.
These relations manifest themselves in underlying structures for the respective
function $G^{[i]}$ ($i = q,g$)~\cite{%
Ravindran:2004mb,Moch:2005tm,Ravindran:2006cg}
such that one can decompose the resummation coefficients $G^{[i]}$ in the massless case
according to
\begin{eqnarray}
\label{eq:gfdec1}
  G_1^{[i]} & = & 2 \left( B_1^{[i]} - \:\delta_{ig} \beta_0 \right) \:\:
     + f_1^{[i]} + \ep \widetilde{G_1^{[i]}}
\, ,
\\
\label{eq:gfdec2}
  G_2^{[i]} & = & 2 \left( B_2^{[i]} - 2 \delta_{ig} \beta_1 \right)
     + f_2^{[i]} + \beta_0 \widetilde{G_1^{[i]}}(\ep\!=\!0)
     + \ep \widetilde{G_2^{[i]}}
\, ,
\\
\label{eq:gfdec3}
  G_3^{[i]} & = & 2 \left( B_3^{[i]} - 3 \delta_{ig} \beta_2 \right)
     + f_3^{[i]} + \beta_1 \widetilde{G_1^{[i]}}(\ep\!=\!0)
     + \beta_0 \Big( \widetilde{G_2^{[i]}}(\ep\!=\!0)
     - \beta_0\widetilde{\widetilde{G_1^{[i]}}}(\ep\!=\!0) \Big)
     + \ep \widetilde{G}_3^{[i]}
\, ,
\quad
\end{eqnarray}
where $i = q,g$ and
\begin{equation}
  \widetilde{F} \:\: = \:\: \ep^{-1}
  \left[ \, F - F (\ep\!=\!0) \, \right]
\, .
\end{equation}
Here (and only here), the functions $B_n^{[i]}$ (not to be confused with the ones given in
Eqs.~(\ref{eq:B1})--(\ref{eq:B3})) denote the coefficients of term
with $\delta(1-x)$ in the $n$-loop diagonal $\MSbar$ splitting
functions $P_{ii}^{(n-1)}$~\cite{Moch:2004pa,Vogt:2004mw}, while the
universal functions $f_n^{[i]}$ exhibit the same maximally
non-Abelian color structure as the
$A_n^{[i]}$~\cite{Korchemsky:1989si} up to the factor $\ca/\cf$,
i.e., $f_i^{[g]} = \ca/\cf \: f_i^{[q]}$, see
Ref.~\cite{Moch:2005tm} for details. With the help of
Eqs.~(\ref{eq:gfdec1})--(\ref{eq:gfdec3}) one could translate the
exponentiated form factor in Eq.~(\ref{eq:massive-resummedff}) for
heavy quarks immediately e.g. to the case of gluinos ${\tilde g}$
with mass $m_{\tilde g}$, all resummation coefficients up to three
loops being known.

For completeness, let us finally mention also the functions $H$ and $S$
of Eq.~(\ref{eq:res-FF}) as well as $B$ and $h$ from Eq.~(\ref{eq:resummedff}).
The function $H$ is already known through
two loops from Refs.~\cite{Korchemsky:1992xv,Gardi:2005yi}, while
the function $S$ was evaluated in Refs.~\cite{%
Corcella:2003ib,Cacciari:2002re} to one loop.
Using Eq.~(\ref{eq:res-FF-fin}) and matching it to the fixed-order
calculation for ${\cal F}_2$ we can extract in particular
the two-loop coefficient $S_2$ from the term $\as^2/\ep L^0$.
The explicit results for $H(\ar)$
of the massive form factor in Eq.~(\ref{eq:res-FF-fin}) read,
\begin{eqnarray}
\label{eq:H1}
  H_1 & = &
%%START
%%L %%texH1 =
          -4 \* \cf
%%;
%%STOP
\, ,
\\
\label{eq:H2}
  H_2 & = &
%%START
%%L %%texH2 =
         \cf \* \ca \* \biggl(
         {220\over 27} + 8 \* \z2 - 36 \* \z3
         \biggr)
         + \nf \* \cf \* {8 \over 27}
%%;
%%STOP
\, ,
\end{eqnarray}
and for $S(\ar)$
\begin{eqnarray}
\label{eq:S1}
  S_1 & = &
%%START
%%L %%texS1 =
           - 4 \* \cf
%%;
%%STOP
\, ,
\\
\label{eq:S2}
  S_2 & = &
%%START
%%L %%texS2 =
         \cf \* \ca \* \biggl(
         - {1396 \over 27} + 8 \* \z2 + 20 \* \z3
         \biggr)
         + \nf \* \cf \* {232 \over 27}
%%;
%%STOP
\, .
\end{eqnarray}
In order to have a self-contained presentation, we also give
the perturbative expansions of the coefficients $B(\ar)$,
\begin{eqnarray}
\label{eq:B1}
  B_1 & = &
%%START
%%L %%texB1 =
  - 3 \* \cf
%%;
%%STOP
\, ,
\\
\label{eq:B2}
  B_2 & = &
%%START
%%L %%texB2 =
    \cf^2 \* \biggl(
    - {3 \over 2}
    + 12 \* \z2
    - 24 \* \z3 \biggr)
  + \cf \* \ca \* \biggl(
    - {3155 \over 54}
    + {44 \over 3} \* \z2
    + 40 \* \z3 \biggr)
\nonumber\\
& &\mbox{}
  + \cf \* \nf \* \biggl(
      {247 \over 27}
    - {8 \over 3} \* \z2 \biggr)
%%;
%%STOP
\, ,
\\
\label{eq:B3}
  B_3 & = &
%%START
%%L %%texB3 =
    \cf^3  \* \biggl(
    - {29 \over 2}
    - 18 \* \z2
    - 68 \* \z3
    - {288 \over 5} \* \z2^2
    + 32 \* \z2 \* \z3
    + 240 \* \z5 \biggr)
\nonumber\\
& &\mbox{}
  + \ca \* \cf^2  \*  \biggl(
    - 46
    + 287 \* \z2
    - {712 \over 3} \* \z3
    - {272 \over 5} \* \z2^2
    - 16 \* \z2 \* \z3
    - 120 \* \z5 \biggr)
\nonumber\\
& &\mbox{}
  + \ca^2 \* \cf  \*  \biggl(
    - {599375 \over 729}
    + {32126 \over 81} \* \z2
    + {21032 \over 27} \* \z3
    - {652 \over 15} \* \z2^2
    - {176 \over 3} \* \z2 \* \z3
    - 232 \* \z5 \biggr)
\nonumber\\
& &\mbox{}
  + \cf^2 \* \nf  \*  \biggl(
      {5501 \over 54}
    - 50 \* \z2
    + {32 \over 9} \* \z3 \biggr)
  + \cf \* \nf^2  \*  \biggl(
    - {8714 \over 729}
    + {232 \over 27} \* \z2
    - {32 \over 27} \* \z3 \biggr)
\nonumber\\
& &\mbox{}
  + \ca \* \cf \* \nf  \*  \biggl(
      {160906 \over 729}
    - {9920 \over 81} \* \z2
    - {776 \over 9} \* \z3
    + {208 \over 15} \* \z2^2 \biggr)
%%;
%%STOP
\, ,
\end{eqnarray}

and for $h(\ar)$
\begin{eqnarray}
\label{eq:h1}
  h_1 & = &
%%START
%%L %%texh1 =
       - 3 \* \cf
       + \ep \* \cf  \*  (  - 16 + 2 \* \z2 )
       + \ep^2 \* \cf  \*  \biggl(  - 32 + 3 \* \z2 + {28 \over 3} \* \z3 \biggr)
       + \ep^3 \* \cf  \*  \biggl(  - 64 + 8 \* \z2
\\
& &\mbox{}
         + 14 \* \z3 + {47 \over 10} \* \z2^2 \biggr)
       + \ep^4 \* \cf  \*  \biggl(  - 128 + 16 \* \z2 + {112 \over 3} \* \z3 + {141 \over 20} \* \z2^2
         - {14 \over 3} \* \z2 \* \z3 + {124 \over 5} \* \z5 \biggr)
\nonumber\\
& &\mbox{}
       + \ep^5 \* \cf  \*  \biggl(  - 256 + 32 \* \z2 + {224 \over 3} \* \z3
         + {94 \over 5} \* \z2^2 - 7 \* \z2 \* \z3 + {186 \over 5} \* \z5 + {949 \over 140} \* \z2^3
         - {98 \over 9} \* \z3^2 \biggr)
%%;
%%STOP
\nonumber \, ,
\\
\label{eq:h2}
  h_2 & = &
%%START
%%L %%texh2 =
         \cf^2  \*  \biggl(  - {3 \over 2} + 12 \* \z2 - 24 \* \z3 \biggr)
       + \cf \* \ca  \*  \biggl(  - {215 \over 6} - {88 \over 3} \* \z2 + 12 \* \z3 \biggr)
       + \nf \* \cf  \*  \biggl( {19 \over 3} + {16 \over 3} \* \z2 \biggr)
\\
& &\mbox{}
       + \ep \* \cf^2  \*  \biggl(  - {1 \over 2} + 116 \* \z2 - 120 \* \z3 - {176 \over 5} \* \z2^2 \biggr)
       + \ep \* \cf \* \ca  \*  \biggl(  - {70165 \over 162} - {575 \over 9} \* \z2 + {520 \over 3} \* \z3
\nonumber\\
& &\mbox{}
         + {176 \over 5} \* \z2^2 \biggr)
       + \ep \* \nf \* \cf  \*  \biggl( {5813 \over 81} + {74 \over 9} \* \z2 - {16 \over 3} \* \z3 \biggr)
       + \ep^2 \* \cf^2  \*  \biggl( {109 \over 4} + 437 \* \z2 - 736 \* \z3 - {432 \over 5} \* \z2^2
\nonumber\\
& &\mbox{}
          + 112 \* \z2 \* \z3 - 48 \* \z5 \biggr)
       + \ep^2 \* \cf \* \ca  \*  \biggl(  - {1547797 \over 972} - {7297 \over 27} \* \z2 + {24958 \over 27} \* \z3
         + {653 \over 6} \* \z2^2
         - {356 \over 3} \* \z2 \* \z3
\nonumber\\
& &\mbox{}
         + 204 \* \z5 \biggr)
       + \ep^2 \* \nf \* \cf  \*  \biggl( {129389 \over 486} + {850 \over 27} \* \z2 - {1204 \over 27} \* \z3
         - {7 \over 3} \* \z2^2 \biggr)
       + \ep^3 \* \cf^2  \*  \biggl( {1287 \over 8} + {2991 \over 2} \* \z2
\nonumber\\
& &\mbox{}
         - 3614 \* \z3
         - 508 \* \z2^2 + 104 \* \z2 \* \z3 - 72 \* \z5 + {6864 \over 35} \* \z2^3 + 1072 \* \z3^2 \biggr)
       + \ep^3 \* \cf \* \ca  \*  \biggl(  - {31174909 \over 5832}
\nonumber\\
& &\mbox{}
         - {155701 \over 162} \* \z2 + {308810 \over 81} \* \z3
         + {100907 \over 180} \* \z2^2 - {478 \over 3} \* \z2 \* \z3 + 840 \* \z5
         - {1618 \over 35} \* \z2^3 - {2276 \over 3} \* \z3^2 \biggr)
\nonumber\\
& &\mbox{}
       + \ep^3 \* \nf \* \cf  \*  \biggl( {2628821 \over 2916} + {8405 \over 81} \* \z2 - {16340 \over 81} \* \z3
         - {1873 \over 90} \* \z2^2
         - {44 \over 3} \* \z2 \* \z3  - 48 \* \z5 \biggr)
%%;
%%STOP
\nonumber \, ,
\\
\label{eq:h3}
  h_3 & = &
%%START
%%L %%texh3 =
         \cf^3  \*  \biggl(  - {29 \over 2}
         - 18 \* \z2 - 68 \* \z3
         - {288 \over 5} \* \z2^2 + 32 \* \z2 \* \z3 + 240 \* \z5 \biggr)
       + \cf^2 \* \ca  \*  \biggl( - {94 \over 3}
\\
& &\mbox{}
         + {1235 \over 3} \* \z2 - {2296 \over 3} \* \z3 + {856 \over 15} \* \z2^2
         - 16 \* \z2 \* \z3 - 120 \* \z5  \biggr)
       + \cf \* \ca^2  \*  \biggl(  - {16540 \over 27} - {22286 \over 27} \* \z2
\nonumber\\
& &\mbox{}
         + {1544 \over 3} \* \z3 + {1592 \over 15} \* \z2^2 - 40 \* \z5 \biggr)
       + \nf \* \cf^2  \*  \biggl( {239 \over 6} - {146 \over 3} \* \z2 + {400 \over 3} \* \z3 - {208 \over 15} \* \z2^2 \biggr)
\nonumber\\
& &\mbox{}
       + \nf \* \cf \* \ca  \*  \biggl( {5516 \over 27}
         + {7216 \over 27} \* \z2 - {224 \over 3} \* \z3
         - {296 \over 15} \* \z2^2 \biggr)
       + \nf^2 \* \cf  \*  \biggl(  - {406 \over 27} - {536 \over 27} \* \z2 \biggr)
%%;
%%STOP
\nonumber \, ,
\end{eqnarray}
where the function $B$ is known to three loops from
Refs.~\cite{Sterman:1987aj,Catani:1989ne,Moch:2002sn,Moch:2005ba},
while the function $h$ has been derived by matching
Eq.~(\ref{eq:resummedff}) to the respective fixed-order calculation
starting from the single pole terms $\as^n/\ep$.

%
% -----------------------------------------------------------------------------
%

\section{Applications}
\label{sec:applications}

Here we want to demonstrate how the previous considerations can be applied
to derive the structure of the singularities and all large Sudakov logarithms
in higher order QCD corrections to partonic scattering processes.
Let us start with the general $2 \to n$ scattering processes of partons $p_i$
in Eq.~(\ref{eq:QCDscattering}) and consider Eq.~(\ref{eq:Mm-M0})
in a perturbative expansion in $\as$.
We want to present the explicit relations between corresponding amplitudes
with and without parton masses $\{m_i\}$, in our notation
$| {\cal M}_{{\rm p},\{m_i\}} \rangle $
and
$| {\cal M}_{{\rm p},\{m_i=0\}} \rangle $.
Throughout this Section, we consider (ultraviolet) renormalized quantities
and we define
\begin{equation}
\label{eq:M-exp}
| {\cal M}_{{\rm p}} \rangle
\, = \,
\sum\limits_{i=0}^{\infty} \left(\ar \right)^i\,
| {\cal M}_{{\rm p}}^{(i)} \rangle
\, ,
\end{equation}
and any overall powers of $\ar$ typical say, for jet cross-sections at
hadron colliders, have been absorbed in the notation.
We can then express Eq.~(\ref{eq:Mm-M0}) for a general process~(\ref{eq:QCDscattering})
in an expansion to second order in
$\as$ as,
\begin{eqnarray}
  \label{eq:2->n-poles-0}
  | {\cal M}_{{\rm p},\{m_i\}}^{(0)} \rangle
  & = &
  | {\cal M}_{{\rm p},\{m_i=0\}}^{(0)} \rangle
\, ,
\\[1ex]
  \label{eq:2->n-poles-1}
  | {\cal M}_{{\rm p},\{m_i\}}^{(1)} \rangle
  & = &
  {1 \over 2}\,
  \sum\limits_{i\in\ \{{\rm all}\ {\rm legs}\}}\,
  Z^{(1)}_{[i]}
  | {\cal M}_{{\rm p},\{m_i=0\}}^{(0)} \rangle
  +
  | {\cal M}_{{\rm p},\{m_i=0\}}^{(1)} \rangle
\, ,
\\
  \label{eq:2->n-poles-2}
  | {\cal M}_{{\rm p},\{m_i\}}^{(2)} \rangle
  & = &
  {1 \over 2}\,
  \sum\limits_{i\in\ \{{\rm all}\ {\rm legs}\}}\,
\left(
  Z^{(2)}_{[i]}
  -
  {1 \over 4}\,
  \left( Z^{(1)}_{[i]} \right)^2
\right)
  | {\cal M}_{{\rm p},\{m_i=0\}}^{(0)} \rangle
\\
& &\mbox{}
  +
  {1 \over 2}\,
  \sum\limits_{i\in\ \{{\rm all}\ {\rm legs}\}}\,
  Z^{(1)}_{[i]}
  | {\cal M}_{{\rm p},\{m_i=0\}}^{(1)} \rangle
  +
  | {\cal M}_{{\rm p},\{m_i=0\}}^{(2)} \rangle
\, ,
\nonumber
\end{eqnarray}
which holds in the small mass limit up to terms suppressed with the parton masses $m_i^2$.
Of course, in the case of massless external lines the
respective higher order corrections to the $Z$-factors
in Eqs.~(\ref{eq:2->n-poles-1}), (\ref{eq:2->n-poles-2}) mostly vanish.
Also recall that the amplitude $| {\cal M}_{\rm p} \rangle $
is a vector in color space whereas the $Z$-factors from Eq.~(\ref{eq:Z}) are in this respect
simply functions.
Non-trivial color dependence of singularities on the other hand typically arises from soft gluon exchange
and therefore carries over directly from underlying
massless hard scattering amplitude $| {\cal M}_{{\rm p},\{m_i=0\}} \rangle $.
Finally, it has been emphasized already in the previous discussions, that
Eqs.~(\ref{eq:2->n-poles-0})--(\ref{eq:2->n-poles-2}) require
to organize the contributions to the massive amplitude $| {\cal M}_{{\rm p},\{m_i\}} \rangle$
in terms of flavor classes, i.e. whether or not the heavy parton lines are external.
An analogous distinction holds for the gluon factor $Z^{(m\vert0)}_{[g]}$ when
heavy quarks are included for instance as self-energy corrections to the
external gluons, see Eqs.~(\ref{eq:IoneTwo}), (\ref{eq:fg1}) below.
The explicit results for $Z^{(1)}_{[q]}$ and $Z^{(2)}_{[q]}$
in Eqs.~(\ref{eq:Zfactor-1loop}), (\ref{eq:Zfactor-2loop}) hold for the cases $ll$, $hl$.

In the light of Eqs.~(\ref{eq:2->n-poles-0})--(\ref{eq:2->n-poles-2})
let us briefly come back to the relation between the factor $Z^{(m\vert0)}$ and
the perturbative fragmentation functions~\cite{Mele:1990cw}.
Although this connection may come at first as a surprise,
the two functions are actually intimately related in the context of QCD amplitudes.
First of all, both are process-independent.
Secondly, one may compare both approaches
in a computation of a one-particle inclusive cross-section of a massive parton
based on an amplitude such as Eq.~(\ref{eq:QCDamplitude}).
The result takes the form of a convolution of massless cross-section
times the perturbative fragmentation function. (We refer the reader to the discussion in
Refs.~\cite{Melnikov:2004bm,Catani:1999ss} for complete details on this point).
Alternatively, we can use Eq.~(\ref{eq:Mm-M0}) to relate the massive amplitude
to the massive one.
As is clear e.g. from the perturbative expansion in
Eqs.~(\ref{eq:2->n-poles-0})--(\ref{eq:2->n-poles-2})
the proportionality factor between
$| {\cal M}_{{\rm p},\{m_i\}} \rangle $
and
$| {\cal M}_{{\rm p},\{m_i=0\}} \rangle $
is independent of the kinematics and is also not affected by the
subsequent phase-space integration.
Furthermore, this holds separately for virtual and
the corresponding real radiation contributions.
Thus, our simple direct relation between massive amplitudes and their
massless counterpart in Eq.~(\ref{eq:Mm-M0}) represents the appropriate
generalization of the formalism of Mele and Nason~\cite{Mele:1990cw}
at the amplitude level.

In an equivalent formulation, we can also consider the perturbative expansion of Eq.~(\ref{eq:QCDfacamplitude}).
To that end, we repeat the decomposition of the amplitude $| {\cal M}_{{\rm p},\{m_i\}} \rangle $
from Eqs.~(\ref{eq:2->n-poles-0})--(\ref{eq:2->n-poles-2})
up to two loops in terms of products of the functions ${\cal J}^{\rm[p]}$, ${\cal S}^{\rm[p]}$
and ${\cal H}^{\rm[p]}$.
\begin{eqnarray}
  \label{eq:2->n-polesJSH-0}
  | {\cal M}_{{\rm p},\{m_i\}}^{(0)} \rangle
  & = &
  | {\cal H}_{\rm p}^{(0)} \rangle
\, ,
\\[1ex]
  \label{eq:2->n-polesJSH-1}
  | {\cal M}_{{\rm p},\{m_i\}}^{(1)} \rangle
  & = &
  {1 \over 2}\,
  \sum\limits_{i\in\ \{{\rm all}\ {\rm legs}\}}\,
  {\cal F}^{[i]}_1
  | {\cal H}_{\rm p}^{(0)} \rangle
  +
  {\cal S}^{\rm [p]}_1
  | {\cal H}_{\rm p}^{(0)} \rangle
  +
  | {\cal H}_{\rm p}^{(1)} \rangle
\, ,
\\
  \label{eq:2->n-polesJSH-2}
  | {\cal M}_{{\rm p},\{m_i\}}^{(2)} \rangle
  & = &
  {1 \over 2}\,
  \sum\limits_{i\in\ \{{\rm all}\ {\rm legs}\}}\,
\left(
  {\cal F}^{[i]}_2
  -
  {1 \over 4}\,
  \left( {\cal F}^{[i]}_1 \right)^2
  +
  {1 \over 2}\,
  {\cal F}^{[i]}_1
  {\cal S}^{\rm [p]}_1
\right)
  | {\cal H}_{\rm p}^{(0)} \rangle
\\
& &\mbox{}
  +
  {1 \over 2}\,
  \sum\limits_{i\in\ \{{\rm all}\ {\rm legs}\}}\,
  {\cal F}^{[i]}_1
  | {\cal H}_{\rm p}^{(1)} \rangle
  +
  {\cal S}^{\rm [p]}_2
  | {\cal H}_{\rm p}^{(0)} \rangle
  +
  {\cal S}^{\rm [p]}_1
  | {\cal H}_{\rm p}^{(1)} \rangle
  +
  | {\cal H}_{\rm p}^{(2)} \rangle
\, ,
\nonumber
\end{eqnarray}
where the perturbative expansions of ${\cal S}^{\rm [p]}$ and $| {\cal H}_{\rm p} \rangle$
are defined analogous to Eq.~(\ref{eq:M-exp}).
Of course, the same qualifications from Section~\ref{sec:qcdffactor} about the distinct
flavor classes contributing to the massive form factor ${\cal F}$ also apply here.
Now, in the factorization ansatz of Eq.~(\ref{eq:QCDfacamplitude}) the function $| {\cal H}_{\rm p} \rangle$
is a vector and ${\cal S}^{\rm [p]}$ is a matrix in color space.
Thus, their products in Eqs.~(\ref{eq:2->n-polesJSH-1}), (\ref{eq:2->n-polesJSH-2})
are in the sense of matrix multiplication
and all dependence on singular color correlations rests entirely in the function ${\cal S}^{\rm [p]}$.

As we remarked above (and as is well known in the literature~\cite{Contopanagos:1997nh,Sterman:2002qn,MertAybat:2006mz})
the matrix ${\cal S}^{\rm[p]}$ is subject of a renormalization group equation
which allows for an all-order exponentiation of the soft contributions.
The solution for ${\cal S}^{\rm[p]}$ results in a path-ordered exponential
due to mixing of the color structures under soft gluon exchange,
\begin{eqnarray}
\label{eq:Sfunction}
{\cal S}^{\rm[p]}\left(\{ k_i \},{Q^2 \over \mu^2},\as(\mu^2),\ep \right) & = &
{\rm P}\, {\rm exp}
\left[
  -{1 \over 2}\int\limits_{0}^{Q^2} {d k^2 \over k^2}
  \Gamma^{\rm[p]}\left({\bar a}(k^2 ,\ep) \right)
\right] \, ,
\end{eqnarray}
where ${\rm P}$ denotes the path ordering.
Here, $\Gamma^{\rm[p]}$ is the so-called soft anomalous dimension, which is a matrix
in the space of color tensors (see Eqs.~(\ref{eq:QCDamplitude}), (\ref{eq:QCDfacamplitude})).
Of course, the running coupling $\bar \as$ in the argument of $\Gamma^{\rm[p]}$ is to be taken in $d$ dimensions.
For $2 \to n$ hard scattering processes with massless partons $\Gamma^{\rm[p]}$ is currently known up to two
loops~\cite{Korchemskaya:1994qp,Jantzen:2005az,MertAybat:2006wq,MertAybat:2006mz} and
to one loop results for reactions with massive partons~\cite{Kidonakis:1997gm,Laenen:1998kp}.
In the latter case, one can show in particular, that the soft anomalous dimension $\Gamma^{\rm[p]}$
has a smooth limit for $m \to 0$.

To summarize, we have given in
Eqs.~(\ref{eq:2->n-poles-0})--(\ref{eq:2->n-poles-2}) and
(\ref{eq:2->n-polesJSH-0})--(\ref{eq:2->n-polesJSH-2}) two
equivalent formulations. Both allow to obtain all large logarithms
of Sudakov type together with the dimensionally regulated soft
poles in $\ep$ and any given QCD amplitude for $2 \to n$ scattering
with parton masses $\{m_i\}$ can be constructed by either method.
In particular Eq.~(\ref{eq:Sfunction}) can be used to derive
explicit expressions for ${\cal S}^{\rm [p]}_1$ and ${\cal S}^{\rm
[p]}_2$ in
Eqs.~(\ref{eq:2->n-polesJSH-0})--(\ref{eq:2->n-polesJSH-2}) in
terms of the perturbative expansion for the soft anomalous
dimension $\Gamma^{\rm[p]}$. Most of the other ingredients are
explicitly presented in this paper.

Next, let us discuss the consistency of Eq.~(\ref{eq:Mm-M0}) with the
results of Ref.~\cite{Catani:2000ef}.
In that reference the structure of both soft and collinear singularities
for any one-loop amplitude was presented for arbitrary values of parton masses.
In the approach of Ref.~\cite{Catani:2000ef}
any one-loop $n$-parton amplitude can be written as:
\begin{eqnarray}
\label{eq:Ione}
| {\cal M}^{(1)}_{\rm p} \rangle &=&  I_n^{(m)}(\ep,\mu^2,\{m_i^2\})\, | {\cal
M}^{(0)}_{\rm p} \rangle + | {\cal M}^{(1),{\rm fin}}_{\rm p}
\rangle
\, ,
\end{eqnarray}
where $| {\cal M}^{(0)}_{\rm p} \rangle$ is the Born amplitude for
the process under consideration. The amplitude $| {\cal
M}^{(1),{\rm fin}}_{\rm p} \rangle$ contains only one-loop
corrections which are finite in the limits $m_i\to 0$ and $\epsilon
\to 0$. In the following we will adapt the results of
Ref.~\cite{Catani:2000ef} to the $\MSbar$ coupling evaluated at a
renormalization scale $\mu$. We will also assume conventional
dimensional regularization for simplicity. In the small mass limit
the operator $I_n^{(m)}$ then takes the form (recall
that all non-vanishing masses $m_j$ are assumed to have a common
value $m$):
\begin{eqnarray}
\label{eq:IoneOp}
I_n^{(m)}(\ep,\mu^2,\{m_i^2\})
&=&
{\exp(\epsilon\gamma_{\rm E}) \over \Gamma(1-\epsilon) }
\Bigg\{ \sum\limits_{j\neq k=1}^{n}
T_j\cdot T_k~
{\cal V}_{jk}(s_{jk};m_j,m_k;\epsilon)
- \sum\limits_{j=1}^n
\Gamma_j(\mu,m_j;\epsilon) +\dots\Bigg\} \, ,
\qquad
\end{eqnarray}
where $T_k$ are the generators of the gauge group and $s_{jk}$ the kinematical
invariants.
The dots denote mass-independent terms and
the functions ${\cal V}_{jk}$ are associated to pairs of external partons.
One has three possible combinations in each pair of partons
with two, one or none of them being massive.
Thus, three separate functions ${\cal V}_{jk}$ are needed for these three cases.
Similarly, the functions $\Gamma_j$ are different, depending on whether the
parton $j$ is massive or massless, i.e. quark, gluon, gluino and so on.

For the sake of comparison with the $Z$-factor in Eq.~(\ref{eq:Z}) we
write ${\cal V}_{jk}$ and $\Gamma_j$ as
\begin{eqnarray}
\label{eq:VtoDelta} {\cal V}^{({\rm 2\ massive\
partons})}_{jk} &=& 2\, \Delta_V + {\cal V}^{(0)}_{jk} \,
\\
{\cal V}^{({\rm 1\ massive\ parton})}_{jk} &=&
\Delta_V + {\cal V}^{(0)}_{jk} \,
\\
\Gamma^{(m)}_{q} &=& \Delta_q + \Gamma^{(0)}_{q}
\,
\\
\Gamma^{(m)}_{g} &=& \nh\, \Delta_g + \Gamma^{(0)}_{g}
\, ,
\end{eqnarray}
where the functions $\Delta_V,\Delta_q$ and $\Delta_g$ are
independent of the invariants $s_{jk}$, i.e. they are the same for
each external parton (or pair of external partons)%
\footnote{The superscripts $(m)$ and $(0)$ refer to
quantities evaluated in the massive, respectively massless case.}.
Therefore one can apply the color algebra to express the sum over
the products of color generators multiplying these functions
directly in terms of the corresponding Casimir operators (see
Ref.~\cite{Catani:2000ef}). In this way, all process dependent
factors are separated into functions that are independent of the
mass. All mass dependence on the other hand enters only in a
process-independent way. Combining the above results
one gets
\begin{eqnarray}
\label{eq:IoneTwo} I_n^{(m)}(\ep,\mu^2,m^2) =  I_n^{(0)}(\ep,\mu^2)
+ \sum\limits_{j=1}^\nh\, f_q(\ep,\mu^2,m^2) +
\sum\limits_{j=1}^\ng \nh f_g(\ep,\mu^2,m^2) \, ,
\end{eqnarray}
where $I_n^{(0)}$ is the appropriate operator for purely massless amplitudes~\cite{Catani:1998bh}
evaluated for $\nf = \nl + \nh$ light flavors.
The function $f_q$ is given by one half of the function $Z^{(1)}_{[q]}$
(and of course restricted to constant terms at order $\ep^0$)
presented in Eq.~(\ref{eq:Zfactor-1loop}).
For the function $f_g$ we find
\begin{eqnarray}
\label{eq:fg1}
f_g(\ep,\mu^2,m^2) = -{1 \over 3} \Biggl({1 \over \ep} + \ln\left({\mu^2 \over m^2}\right) \Biggr)
\, ,
\end{eqnarray}
which, when restricted to constant terms at order
$\ep^0$, is related to the function $Z^{(1)}_{[g]}$ in Eq.~(\ref{eq:Zfactor-gluon})
through $\nh f_g(\ep,\mu^2,m^2) = Z^{(1)}_{[g]}/2$.

Finally, we briefly comment on Abelian gauge theories with fermion masses
like Quantum Electrodynamics (QED).
These provide other prominent examples for the considerations of the present
article.
For instance one arrives at QED (with massive electrons) after the usual identification
of the color factors,
$\,C_F=1$, $C_A=0$ and $T_f=1$ instead of our QCD convention $T_f\, \nf =\nf/2$
There, the complete calculation of the two-loop radiative photonic corrections in QED
to Bhabha scattering in the small mass limit have already been
performed~\cite{Glover:2001ev,Penin:2005kf,Penin:2005eh}.
This included also a complete matching at two loops, i.e. the computation of
the constant terms which are not logarithmically enhanced.
The latter also required the constant terms from the massless,
dimensionally regularized amplitudes of Ref.~\cite{Bern:2000ie}.
An extension of the results of the present article (and the exponentiation in particular)
in this direction is a possibility which we leave for a future publication.

%
% -----------------------------------------------------------------------------
%
\section{Summary}
\label{sec:summary}

In this article we have presented a first discussion of the
singular behavior of on-shell QCD amplitudes with massive particles
beyond one loop. We have performed a systematic study of both, the
soft singularities typically showing up as poles in $\ep$ in
dimensional regularization and the structure of the large Sudakov
(or quasi-collinear) type logarithms of the parton masses, which
become dominant in the high energy limit.
Working in the small mass limit, we have consistently omitted power corrections
in the parton masses.

We have presented in Eqs.~(\ref{eq:QCDfacamplitude}) and (\ref{eq:Mm-M0})
a general framework for the factorization of
$n$-parton amplitudes in QCD which incorporates massive partons.
The factorization formula, which we have organized in terms of flavor
classes, is universal and is valid for any amplitude.
We have emphasized the
strong similarities between scattering amplitudes with massless and
massive partons in the limit where all parton masses are much
smaller than the relevant kinematic invariants of the scattering
process. In this regime, the factorization formula can be used to
directly obtain (apart from vanishing corrections when the masses
tend to zero) the massive amplitude from the corresponding massless
amplitude, without explicitly computing the former.
To that end we have introduced the factor $Z^{(m\vert0)}$
as the building block of the proportionality.
In the case of heavy quarks
we have linked $Z^{(m\vert0)}_{[q]}$ to the virtual corrections
in the formalism of perturbative fragmentation function
thus generalizing the approach of Ref.~\cite{Mele:1990cw} to the level of
amplitudes.
Finally, we have explicitly illustrated the predictive power of the factorization
ansatz for examples from $2 \to n$ scattering
processes in QCD.

Improved insight into the structure underlying the factorization of
amplitudes in the soft and (quasi)-collinear momentum regions have
enabled us to derive an exponential~(\ref{eq:massFFexp}) for the
form factor of heavy quarks. We have used this new result to
predict the fixed-order expansion of the massive form factor to up
three loops and, in comparing massless and massive amplitudes, we
have observed an apparent universality of the respective
resummation coefficients $G$ which we find worth mentioning.
Furthermore, on the basis of Eq.~(\ref{eq:QCDfacamplitude}) and the
exponentiations for the functions ${\cal J}^{\rm[p]}$ and ${\cal
S}^{\rm[p]}$ we have shown how to extend our predictions to the
perturbative expansion of general $n$-parton amplitudes in QCD with
massive partons.

Thus, the results of the present paper such as Eq.~(\ref{eq:Z}) can
be useful to either check explicit evaluations of amplitudes at
higher loops or make predictions to higher orders in perturbation
theory. The material presented can also help to organize
calculations, say at NNLO, in terms of divergent, but analytically
computable, parts and finite remainders that can be integrated
numerically. In the context of general calculations for
differential observables with massive partons at NNLO our
factorization formula may also facilitate the combination of the
respective tree-level and one-loop real emission amplitudes with
the virtual contributions in a process independent manner.

We will return to these issues as well as potential connections to
threshold resummation for processes with massive partons in future
work.

%
% ---------------------------------------------------------------------
%
{\bf{Acknowledgments:}}
We are grateful to L.~Dixon and A.~Vogt for insightful
discussions and careful reading of the manuscript.
We also acknowledge useful discussions with S.~Actis, S.~Catani, J.~Gluza,
Z.~Merebashvili and T.~Riemann.
The Feynman diagrams have been created with the packages {\sc
Axodraw}~\cite{Vermaseren:1994je} and {\sc
Jaxo\-draw}~\cite{Binosi:2003yf}. A.M. thanks the Alexander von Humboldt
Foundation for support and the theory groups at
SLAC and Fermilab for their warm hospitality during the initial
stage of this project.
S.M. acknowledges contract VH-NG-105 of the
Helmholtz Gemeinschaft.
This work is supported in part by the Deutsche Forschungsgemeinschaft in
Sonderforschungs\-be\-reich/Transregio~9
and a preliminary version of this work was reported
at the workshop ${\rm HP}^2$ in Z\"urich, Switzerland, Sep. 06-09, 2006.

\appendix
%
% ---------------------------------------------------------------------
%
\renewcommand{\theequation}{\ref{sec:appA}.\arabic{equation}}
\setcounter{equation}{0}
\section{Appendix}
\label{sec:appA}

Here, we give the complete result for one-loop QCD corrections to order $\ep^2$
to the form factor of a heavy quark at the scale $\mu^2 = m^2$ in terms of
harmonic polylogarithms $H_{m_1,...,m_w}(x)$,
see also Section~\ref{sec:qcdffactor} and Eq.~(\ref{eq:massFFpert}) for definitions.
The variable $x$ with $0 \leq x \leq 1$ for space-like $q^2 = - Q^2<0$ is given by
\begin{equation}
\label{eq:xvar}
x = {{\sqrt{Q^2+4m^2} - \sqrt{Q^2} } \over {\sqrt{Q^2+4m^2} + \sqrt{Q^2}}}
\ .
\end{equation}
For ${\cal F}_1$ we find\footnote{
We thank J.~Gluza for providing us with the integral {\tt SE2l2m} of
Refs.~\cite{Czakon:2004wm,Czakon:2006pa} to order $\ep^3$,
see also {\tt http://www-zeuthen.desy.de/theory/research/bhabha/}.
}
\begin{eqnarray}
\label{eq:F1exact}
  {\cal F}_1 & = &
%%START
%%L %%texF1exact =
       {1 \over \ep} \* \cf \* \biggl\{
          - 2
          + 2 \* \biggl(1 - {1 \over {1-x}} - {1 \over {1+x}}\biggr)\* \H(0)
        \biggr\}
       + \cf \* \biggl\{
          - 4
          + \biggl(3 - {4 \over {1-x}} - {2 \over {1+x}}\biggr) \* \H(0)
\\
& &\mbox{}
          + 2 \* \biggl(1 - {1 \over {1-x}} - {1 \over {1+x}}\biggr) \* (
            \Hh(0,0)
          - 2 \* \Hh(-1,0)
          - \z2
          )
        \biggr\}
       + \ep \* \cf \* \biggl\{
          - 8
          + \biggl(1 - {1 \over {1-x}} - {1 \over {1+x}}\biggr) \* (
            8 \* \H(0)
\nonumber\\
& &\mbox{}
          + 8 \* \Hhh(-1,-1,0)
          - 4 \* \z3
          + 4 \* \H(-1) \* \z2
          - 4 \* \Hhh(-1,0,0)
          - \H(0) \* \z2
          - 4 \* \Hhh(0,-1,0)
          + 2 \* \Hhh(0,0,0)
          )
\nonumber\\
& &\mbox{}
          + \biggl(3 - {4 \over {1-x}} - {2 \over {1+x}}\biggr) \* (
            \Hh(0,0)
          - 2 \* \Hh(-1,0)
          )
          - 2 \* \biggl(2 - {2 \over {1-x}} - {1 \over {1+x}}\biggr) \* \z2
        \biggr\}
       + \ep^2 \* \cf \* \biggl\{
          - 16
\nonumber\\
& &\mbox{}
          - {4 \over 3} \* \biggl(4 - {6 \over {1-x}} - {3 \over {1+x}}\biggr) \* \z3
          + \biggl(3 - {4 \over {1-x}} - {2 \over {1+x}}\biggr) \* \biggl(
            2 \* \H(-1) \* \z2
          + 4 \* \Hhh(-1,-1,0)
          - 2 \* \Hhh(-1,0,0)
\nonumber\\
& &\mbox{}
          - {1 \over 2} \* \H(0) \* \z2
          - 2 \* \Hhh(0,-1,0)
          + \Hhh(0,0,0)
          \biggr)
          + \biggl(1 - {1 \over {1-x}} - {1 \over {1+x}}\biggr) \* \biggl(
          - {14 \over 5} \* \z2^2
          + 8 \* \H(-1) \* \z3
          - 8 \* \Hh(-1,-1) \* \z2
\nonumber\\
& &\mbox{}
          - 16 \* \Hhhh(-1,-1,-1,0)
          + 8 \* \Hhhh(-1,-1,0,0)
          - 16 \* \Hh(-1,0)
          + 2 \* \Hh(-1,0) \* \z2
          + 8 \* \Hhhh(-1,0,-1,0)
          - 4 \* \Hhhh(-1,0,0,0)
          + 16 \* \H(0)
\nonumber\\
& &\mbox{}
          - {14 \over 3} \* \H(0) \* \z3
          + 4 \* \Hh(0,-1) \* \z2
          + 8 \* \Hhhh(0,-1,-1,0)
          - 4 \* \Hhhh(0,-1,0,0)
          + 8 \* \Hh(0,0)
          - \Hh(0,0) \* \z2
          - 4 \* \Hhhh(0,0,-1,0)
\nonumber\\
& &\mbox{}
          + 2 \* \Hhhh(0,0,0,0)
          \biggr)
          - 2 \* \biggl(5 - {4 \over {1-x}} - {4 \over {1+x}}\biggr) \* \z2
        \biggr\}
%%;
%%STOP
\nonumber
\, .
\end{eqnarray}
In Eq.~(\ref{eq:F1exact}) all harmonic polylogarithms
$H_{m_1,...,m_w}(x)$, $m_j = 0,\pm 1$ are understood to be of
argument $x$. For the rest, our notation follows
Ref.~\cite{Remiddi:1999ew} to which the reader is referred for a
detailed discussion.

Next we present the one-loop result for the virtual contribution to the
perturbative fragmentation function to all orders in $\epsilon$
\cite{Melnikov:2004bm}:
\begin{equation}
\label{eq:Diniv-oneloop}
D^{\rm virt}_1(z) \, = \,
\ar\cf\, {2\ep^2-3\ep+2\over (1-2\ep)\ep}\,
\exp(\ep\gamma_{\rm E})\Gamma(\ep)\,
{\left(\mu^2\over m^2\right)}^\ep\, \delta(1-z)
\, .
\end{equation}

As one can easily verify, the expansion of the coefficient
of the delta-function in $\ep$ coincides to all known powers with the factor
$Z^{(1)}_{[q]}$ in Eq.~(\ref{eq:Zfactor-1loop}),
which suggests that Eq.~(\ref{eq:Diniv-oneloop})
is indeed the proper generalization to all orders in $\ep$.

Finally, we discuss the derivation of the one-loop
heavy quark insertion in the tree-level gluon form factor. It is
clear that this diagram coincides with the one loop contribution
to $Z^{(1)}_{[g]}$. We find it particularly convenient to evaluate
this diagram in a physical light-cone gauge. Following the
procedure outlined in Refs.~\cite{Melnikov:2004bm,Mitov:2004du} we
obtain to all orders in $\ep$:
\begin{equation}
\label{eq:Z-gluon-all-eps}
Z^{(1)}_{[g]} \, = \, \ar n_h \,
\left(-{2 \over 3}\right) \, \exp(\ep\gamma_{\rm E})\Gamma(\ep)\,
{\left(\mu^2\over m^2\right)}^\ep \, .
\end{equation}
Upon expansion in $\ep$ Eq.~(\ref{eq:Zfactor-gluon}) is derived.

%
% -----------------------------------------------------------------------------
%
{\footnotesize
%\bibliography{../Bib/refs}
%\bibliographystyle{../Bib/h-elsevier2}

}

\end{document}